\documentclass[preprint2]{aastex}
\usepackage{natbib}
\usepackage{amssymb}
\pdfoutput=1
\newcommand{\hone}{{{\sc H i}~}}

\slugcomment{To appear in the Astronomical Journal}

\shorttitle{\hone Clouds in the M81/M82 Group}
\shortauthors{Chynoweth et al.}

\begin{document}

\title{Neutral Hydrogen Clouds in the M81/M82 Group}

\author{Katie M. Chynoweth\altaffilmark{1}}
\affil{Vanderbilt University, Physics and Astronomy Department, 1807 Station B, 
Nashville, TN 37235}

\author{Glen I. Langston}
\affil{National Radio Astronomy Observatory,
Green Bank, WV 24944}

\author{Min S. Yun}
\affil{University of Massachusetts, 710 North Pleasant Street, 
Amherst, MA 01002}

\author{Felix J. Lockman, K.H.R. Rubin\altaffilmark{2} and 
Sarah A. Scoles\altaffilmark{3}}
\affil{National Radio Astronomy Observatory,
Green Bank, WV 24944}

\altaffiltext{1}{Graduate Summer Research Assistantship Program at NRAO.  
Present address: Vanderbilt University, Physics and Astronomy Department, 1807 Station B, 
Nashville, TN 37235}
\altaffiltext{2}{Research Experience for Undergraduates at NRAO.  
Present address: UCO/Lick Observatory, University of California, Santa Cruz, CA 95064}
\altaffiltext{3}{Research Experience for Undergraduates at NRAO.  
Present address: Cornell Space Sciences Building, Cornell University, Ithaca, NY 14853}

\begin{abstract}
We have observed a  $3^{\circ} \times 3^{\circ}$ area centered on the M81/M82 group of galaxies using the Robert C. Byrd Green Bank Telescope (GBT) in a search for analogs to the High Velocity Clouds (HVCs) of neutral hydrogen found around our galaxy.   The velocity range from -605 to -85 km s$^{-1}$ and 25 to 1970 km s$^{-1}$ was searched for \hone clouds.  Over the inner  $2^{\circ} \times 2^{\circ}$ the 7$\sigma$ detection threshold was $9.6 \times 10^{5} M_{\odot}$. We detect 5 previously unknown \hone clouds associated with the group, as well as numerous associated filamentary \hone structures, all lying in the range $-105 \leq V_{helio} \leq +280$  km s$^{-1}$.  From the small angular distance of the clouds to group members, and the small velocity difference between group members and clouds, we conclude that the clouds are most likely relics of ongoing interactions between galaxies in the group.

\end{abstract}

\keywords{galaxies: clusters: individual(M81) ---
galaxies: interactions --- surveys --- radio lines: galaxies}

\section{Introduction}
\hone clouds are found around some spiral galaxies at velocities not simply related to the rotation of the galactic disk. 
In the Milky Way there are the  high velocity clouds (HVCs) both ionized and neutral  found  in large complexes and as smaller isolated clouds which together cover nearly 40\% of the sky \citep{wvw97, loc02}.  M31 has a similar system  \citep{thilker04}.  These clouds most likely result from a variety of phenomena.  Interactions between galaxies can produce filamentary structures of gas and tidal streams like the Magellanic Stream \citep{put03} and the tidal tails of the 
M81 group \citep{yun93, wal02}.  Some HVCs may be  the remnants of galaxy formation that are currently being accreted \citep{mal04}, or clumps and filaments of \hone associated with faint dwarf galaxies \citep{thilker04, bekhti06, beg06} .  Some clouds may also arise from a ``galactic fountain", wherein hot gas created by supernovae 
expands thermally, cools, and condenses and falls back into the galaxy (see \citet{wvw97} and references therein).  Studies of cloud properties such as mass,  morphology, and kinematics are necessary to distinguish between these possibilities.

Measurement of many physical quantities is dependent on an accurate value for distance. The exact distance of the clouds in our galaxy is very difficult to measure, so determination of their properties has been difficult. Many attempts have been made to measure the distances of clouds in the Milky Way (see, e.g., \citet{sch95,vanw99, bekhti06}). However, these efforts have succeeded only in placing limits on cloud distances. 

Other galaxies may also have these high velocity clouds. 
For extragalactic sources, distances from a \hone cloud to its associated galaxy can be determined to a high degree of accuracy, so their properties may be studied more readily.  
We observed the M81/M82 cluster with the GBT in order to map its neutral hydrogen cloud population.  
The group contains optically bright galaxies M81, M82, NGC 3077, and NGC 2976, which show the remnants of strong interactions \citep{wal02}, as well as over 40 dwarf galaxies. See Table 1 for a summary of galaxies within our field of observation. The M81 group has been the subject of many studies including, e.g., \citet{rob72}, \citet{app81}, \citet{lo79}, \citet{yun93}, and \citet{boy01}. \citet{loc02} also found diffuse neutral hydrogen emission in the direction of the M81 group.  They find locations where the emission is likely due to the interactions of M81/M82 but also find lower level emission away from known galaxies. 

In addition to our velocity range search of -605 to -85 km s$^{-1}$ and 25 to 1970 km s$^{-1}$, we have achieved a lower mass detection threshold than previous studies. Our observations have a mass sensitivity 50-100\% lower than the studies by \citet{app81} and \citet{boy01}, due to our high velocity resolution and the excellent sensitivity of the GBT. Our observations cover a large angular area in order to make a complete study of the \hone cloud population and properties. The beam size of the GBT at L band is well matched to the expected angular size of these objects, so our study provides new insight into their structure. The study of \citet{boy01} using HIJASS data covered a larger angular area ($8^{\circ} \times 8^{\circ}$) than our observations, but with a much smaller velocity range search (-500 to 500 km s$^{-1}$) and a coarser (26 km s$^{-1}$) velocity resolution.

In addition to distinct clumps of hydrogen, galaxies are found to have filamentary structures of \hone gas and tidal streams associated with interactions \citep{yun93, wal02}.  \citet{beg06} also found clumps and filaments of \hone associated with faint dwarf galaxies.  The M81 group is notoriously rich in large-scale \hone filaments, a clear signature of interaction between the galaxies. We include in our study a search for such filamentary structures and smaller scale filaments that might provide additional information about interactions.

\hone clouds are also related to dwarf galaxies. In a study of nine Local Group dwarf galaxies, \citet{bouchard06} found that the concentration of HVCs was markedly enhanced near dwarfs, and that there were \hone clouds near (in projection) most of the dSphs and dIrr/dSphs of the LG, although they were typically offset from the optical center.  \citet{bli00} found that many \hone envelopes of dwarf galaxies in the local group were very similar to extragalactic \hone clouds in mass and size, and although most neutral hydrogen clouds are thought to contain no stars \citep{siegel05}, they found that two previously cataloged clouds actually harbored dwarf galaxies. These results suggest a strong relationship between \hone clouds and dwarf galaxies, although the nature of the connection has yet to be fully elucidated. Perhaps some \hone clouds enshroud low surface-brightness galaxies, or are ejected from dwarf galaxies through other mechanisms.   

\citet{bouchard06} concluded that ram pressure or tidal stripping are not likely to be the cause of the observed offset of \hone clouds from their associated galaxies. An alternative explanation is that the gas could be blown out by star formation.  Recently, \citet{wal02} have shown that episodes of star formation in the dwarf galaxy NGC 3077 have disrupted and ejected the gas outside of the region where the stars are optically identified.  Our observations cover a large angular area around this galaxy (as well as others in the group) to search for these blown out components. In addition, these larger angular regions enable us to distinguish foreground components in our galaxy and smaller angular size components of dwarf galaxies in the group.

In sections \ref{obsec} and \ref{datasec} we summarize the observations and data reduction. In section \ref{interp} we describe methods used to interpret our observations. In section \ref{knownmsec}, mass calculations for known objects in the group are presented. In section \ref{ressec} we detail new detections, including \hone clouds and filaments, and their properties. In section \ref{simsec}, our new detections are compared with a numerical simulation of the M81/M82 
group. Finally, in section \ref{consec} we discuss the implications of our results and future work.

\section{Observations}
\label{obsec}
We observed the M81/M82 group with the 100m Robert C. Byrd Green Bank Telescope (GBT) of the NRAO\footnote{The National Radio Astronomy Observatory (NRAO) is a facility of the National Science Foundation operated under cooperative agreement by Associated Universities, Inc.} on 12-16 June, 2003.  The full group extends over a $40 \times 20$ degree area (right ascension range 7 to 11 hours, declination range $58^{\circ} \lesssim \delta \lesssim 78^{\circ}$), but the brightest galaxies (M81, M82 and NGC3077) are all within a $2^{\circ} \times 2^{\circ}$  region centered on M81, which was where we concentrated our observations.  The  $2^{\circ} \times 2^{\circ}$ region  was observed by moving the telescope in declination and  sampling every $4\arcmin$ at an integration time of 5 seconds per sample.  Strips of constant declination were spaced by  $2.\arcmin7$.   A larger field was then sampled at $4\arcmin$ spacing and an integration time of one second per spectrum.  The data discussed here cover an area 
$3^{\circ} \times 3^{\circ}$ centered on M81 to include NGC 2976 and the extended \hone emission associated the group. 

The typical system temperature for each channel of the dual-polarization receiver was 20 K.  The GBT spectrometer was used to cover a velocity range -3172 $< V_{helio} <$ 7376 km s$^{-1}$.  Spectra were smoothed to an effective 
channel spacing of 1.29  km s$^{-1}$.  Over the inner 4 square degree region, the resultant typical  RMS noise is 18 mK;  outside the central region  the noise in a channel rises to  $\approx51$ mK.  The instrumental parameters are summarized in Table 2, and Table 3 gives a summary of observations. 

\section{Data Reduction}
\label{datasec}

The archived GBT data were reduced in the standard manner using the GBTIDL\footnote{Developed by the National Radio Astronomy Observatory; documentation at http://gbtidl.sourceforge.net} data reduction package.

In order to match our velocity resolution to the expected linewidths of \hone clouds in the group, we smoothed the data to contain 2048 channels with a channel spacing of 24.4 kHz, corresponding to a velocity resolution of 5.2 km s$^{-1}$. 
A reference spectrum for each set of scans was made using the first and last scan of the set. The reference spectrum was then used to perform a  (signal-reference)/reference calibration of each integration of each scan. The calibrated spectra were scaled by the system temperature,  corrected for atmospheric opacity and GBT efficiency. We adopted GBT efficiency equation (1) from \citet{lang07}. We assumed a zenith atmospheric opacity $\tau_{0}$ = 0.009. 
The GBT observations were performed without Doppler tracking.  Within GBTIDL the velocities were shifted to the heliocentric frame. 

After amplitude calibration, the spectral bandpass variation was removed by subtraction of a median filtered baseline with 1300 km s$^{-1}$ width (6.0 MHz width). The median filter baseline subtraction process is summarized here and is described with example figures in \citet{lang07}. The baseline was subtracted for each polarization separately, in order
to remove the Intermediate Frequency (IF) electronics gain variations that are unique to each signal path.   The baseline subtraction process also removes gain and sky variations between the time of observation and the time of observation of the reference scan. The baseline value is computed for each individual channel of the spectrum
by computing the median value of all channels within a range of channels centered on the individual channel.
The median width must be small enough to preserve the gain variations, yet large enough to reject spectral lines within the median range. After computation, the baseline is subtracted from the original spectrum.

Calibrated GBTIDL `keep' files were processed using {\tt idlToSdfits} \footnote{Developed by Glen Langston of NRAO; documentation at http://wiki.gb.nrao.edu/bin/view/Data/IdlToSdfits}. This program is used to flag data contaminated by
radio frequency interference (RFI) and convert the data to the form required for input into the NRAO AIPS\footnote{Developed by the National Radio Astronomy Observatory; documentation at http://www.aoc.nrao.edu/aips} package.
The frequency range observed was relatively free of RFI and less than 0.2 \% of all spectra were adversely affected by RFI. The spectra exhibiting RFI values were identified by tabulation of the RMS noise level in channels free of neutral hydrogen emission. In order to remove RFI from the data, {\tt idlToSdfits} was run on each set of scans, and the resulting RMS noise values were examined. Spectra that showed high values of RMS noise across many channels were flagged and removed. After calibration, the observations were imported into the AIPS package. Observations were gridded using the AIPS task SDIMG, which also averages polarizations.

The beam size was found by fitting a two dimensional Gaussian to the continuum emission from M82. The fitted beam was 10.1\arcmin $\pm$ 0.6\arcmin $\times$ 9.4\arcmin $\pm$ 0.5\arcmin, at a position angle of 54$^{\circ}$  $\pm$ 32$^{\circ}$. In order to convert to Jy/beam, we observed the calibration source 3C286. \citet{ott94} provide a model for the flux density of 3C286 relative to other radio sources.  For 3C286 at 1.41 GHz the flux density is 14.61 $\pm$ 0.94 Jy.  Based upon the ratio of 3C286 to M82 intensities, the flux density of M82 is 7.17 $\pm$ 0.46 Jy. Our value for M82 is consistent with 6.2 $\pm$ 0.2 Jy from \citet{con98}. We determined the calibration from from K to Jy by producing a continuum image of M82 in the same manner as the spectral line images.  The peak M82 brightness temperature (K) was measured relative to laboratory measurement calibration noise diodes. The calibration factor was determined by the ratio of the measured brightness temperature divided by flux density of M82 (Jy). Including all corrections for the GBT efficiency and the mapping process, the scale factor is  0.47 Jy/K.

\section{Data Interpretation}
\label{interp}
For all estimates of \hone masses, we adopt the M81 distance, $D = 3.63 \pm 0.34$ Mpc from \citet{kar04}.  M81 is at a distance between M82 and NGC3077,  3.52 and 3.82 Mpc respectively. 

In their study of \hone clouds in M31, \citet{thilker04} found clouds mostly within 150 km s$^{-1}$ of the systemic velocity of M31.  Filaments were found within 80 km s$^{-1}$ of the galaxy's velocity.  If these values can be interpreted as a prediction for \hone cloud locations, we expect to find clouds with velocities within 150 km s$^{-1}$ relative to the main galaxies in the group. For this paper we use galaxy velocities from \citet{yun99}. For completeness, spectral maps were made in a wide velocity range, from -605 to 1970 km s$^{-1}$.  This extends the search to more than $\pm$ 500 km s$^{-1}$ beyond the systemic velocity of any group galaxy. The 7$\sigma$ detection threshold for \hone mass was 
$9.6 \times 10^{5} M_{\odot}$.  Spectral maps were visually inspected for possible new clouds. Objects that were spatially distinct from group galaxies and tidal streams in a range of velocity were considered to be cloud candidates. 
Positions of candidates were measured by fitting a 2-dimensional Gaussian with the AIPS tasks JMFIT and IMSTAT to the peak emission.  For each \hone cloud candidate, spectra were obtained from ISPEC, and \hone column density maps for the velocity range in which the candidate was detected were made with MOMNT. 

Our search area of 3$^\circ$ $\times$ 3$^\circ$ centered on M81 corresponds to 194 $\times$ 194 kpc$^{2}$.  
Based on studies of the Milky Way and M31, which found clouds anywhere within 1-50 kpc of the center of a galaxy \citep{thilker04, wvw97}, we expected that \hone clouds should be located within 50 kpc of associated galaxies. 
Therefore, we should detect all of the \hone clouds around M81.  For M82, NGC 3077, and NGC 2976, our field of view excludes some of the regions where clouds might be found; as such, our detections for those galaxies will place lower limits on the \hone cloud population.  \citet{thilker04} found that filamentary structures were limited to the inner 30 kpc 
of the galaxy, so we expect to detect all such structures associated with the M81 group.

%
%
The boundaries of regions containing clouds were determined by inspection of the spectral line images. The boundaries of the region averaged into the \hone intensity profile were chosen to separate the individual components from the other galaxies.  However all of the newly identified clouds are in close proximity to the interacting members of the galaxies. For each profile, the region was set using a box approximating the area of neutral hydrogen most clearly associated with the object in question and  not involved in interaction. 

The spectral line image is available on-line\footnote{The spectral line image cube is available at http://www.nrao.edu/astrores/m81}. 

\section{Masses of Known Objects}
\label{knownmsec}
There is some overlap in velocity space between the M81/M82 group and the Milky Way, and over this velocity range we can not discriminate between local  and distant emission. To estimate the total \hone mass of neutral hydrogen for the M81/M82 group we replaced the velocity channels contaminated by Milky Way and local high velocity clouds with interpolated values computed from spectral line observations at -85 and +25 km s$^{-1}$.  

\begin{figure}
\centering
\begin{tabular}{cc}
\includegraphics[width=0.333\linewidth, height=0.333\linewidth]{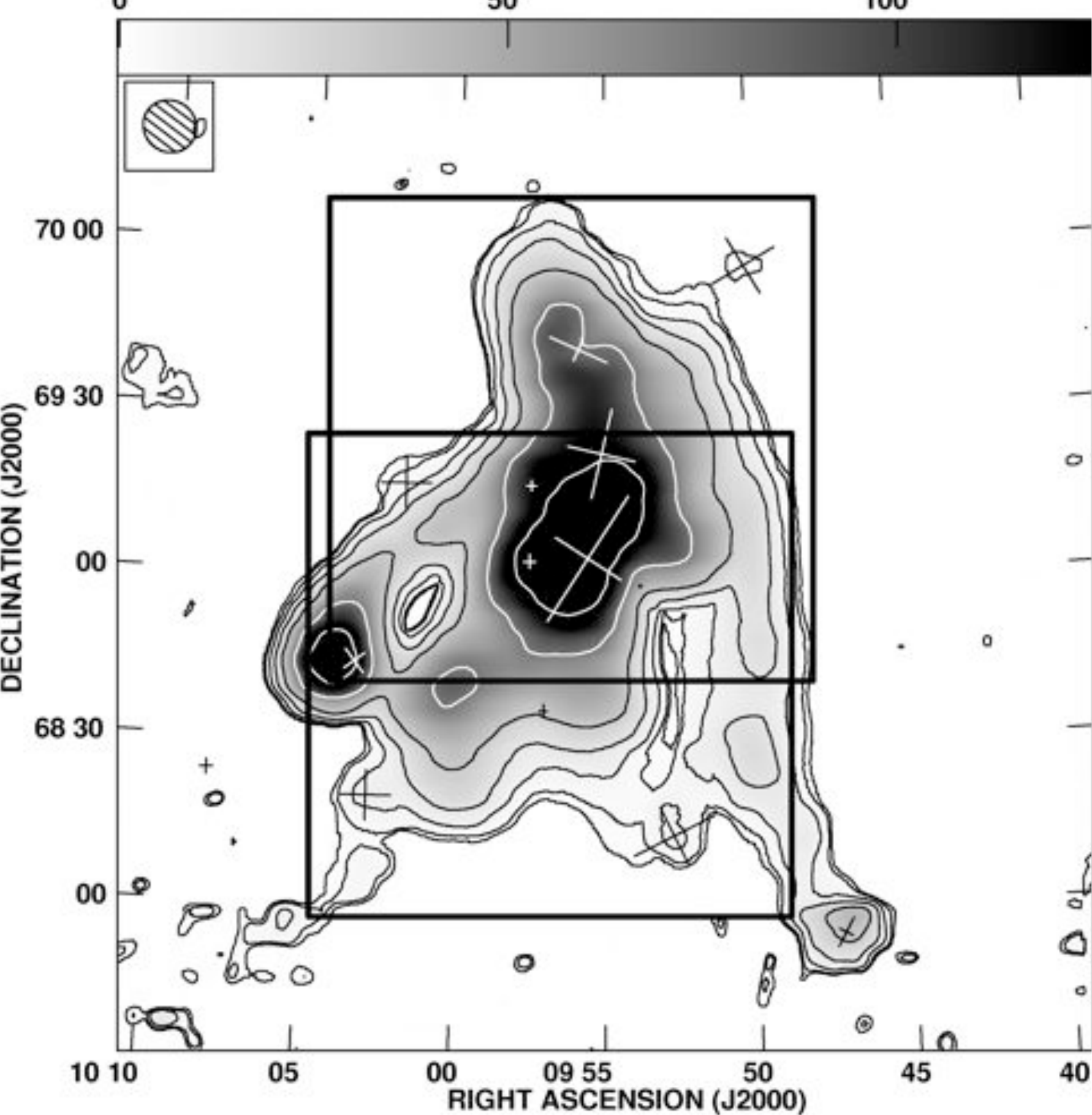} & 
\includegraphics[width=0.333\linewidth, height=0.333\linewidth]{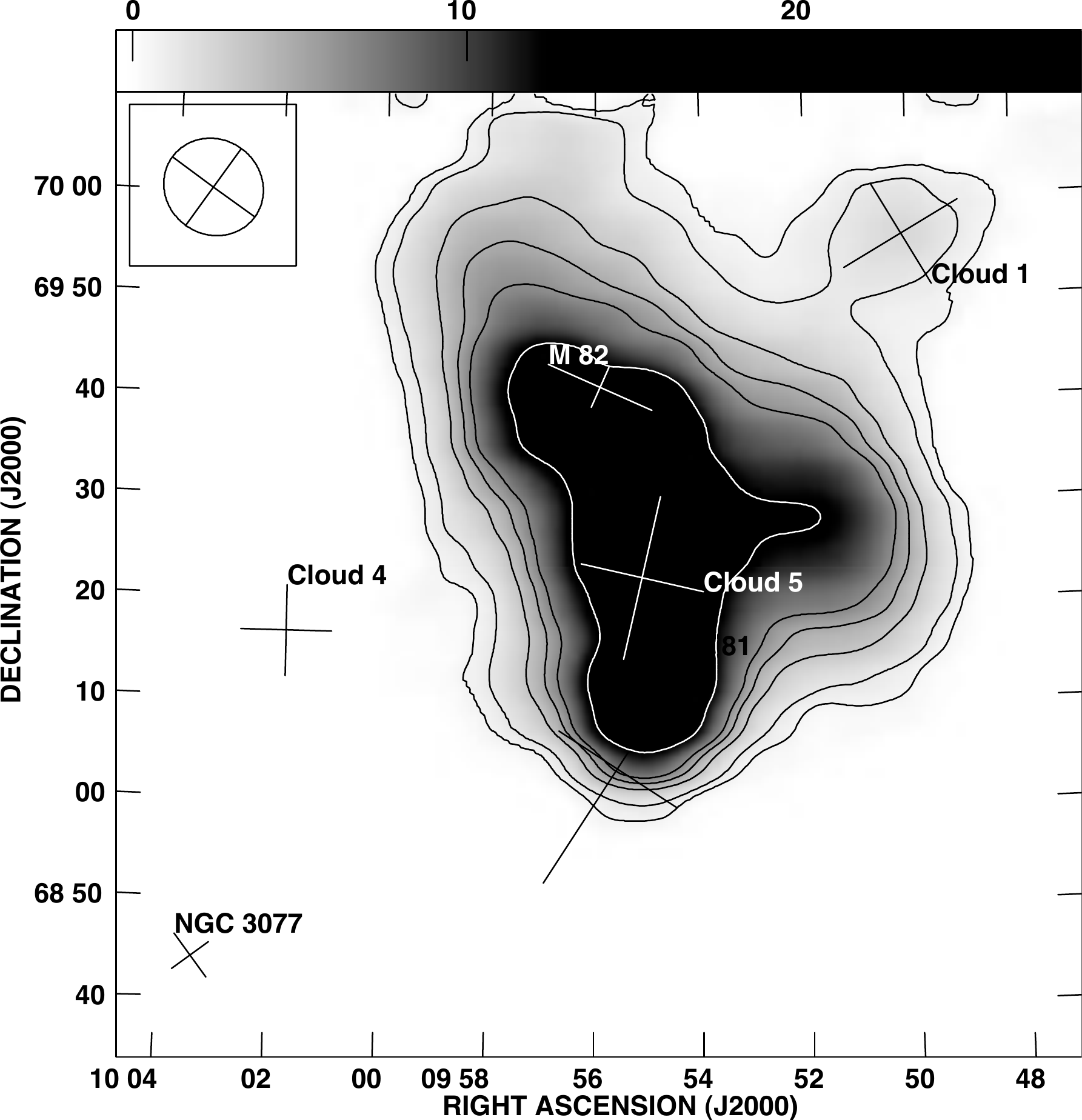}  \\
Field of View (-250 to 340 km s$^{-1}$) & Cloud 1 (160 to 185 km s$^{-1}$)\\
\includegraphics[width=0.333\linewidth, height=0.333\linewidth]{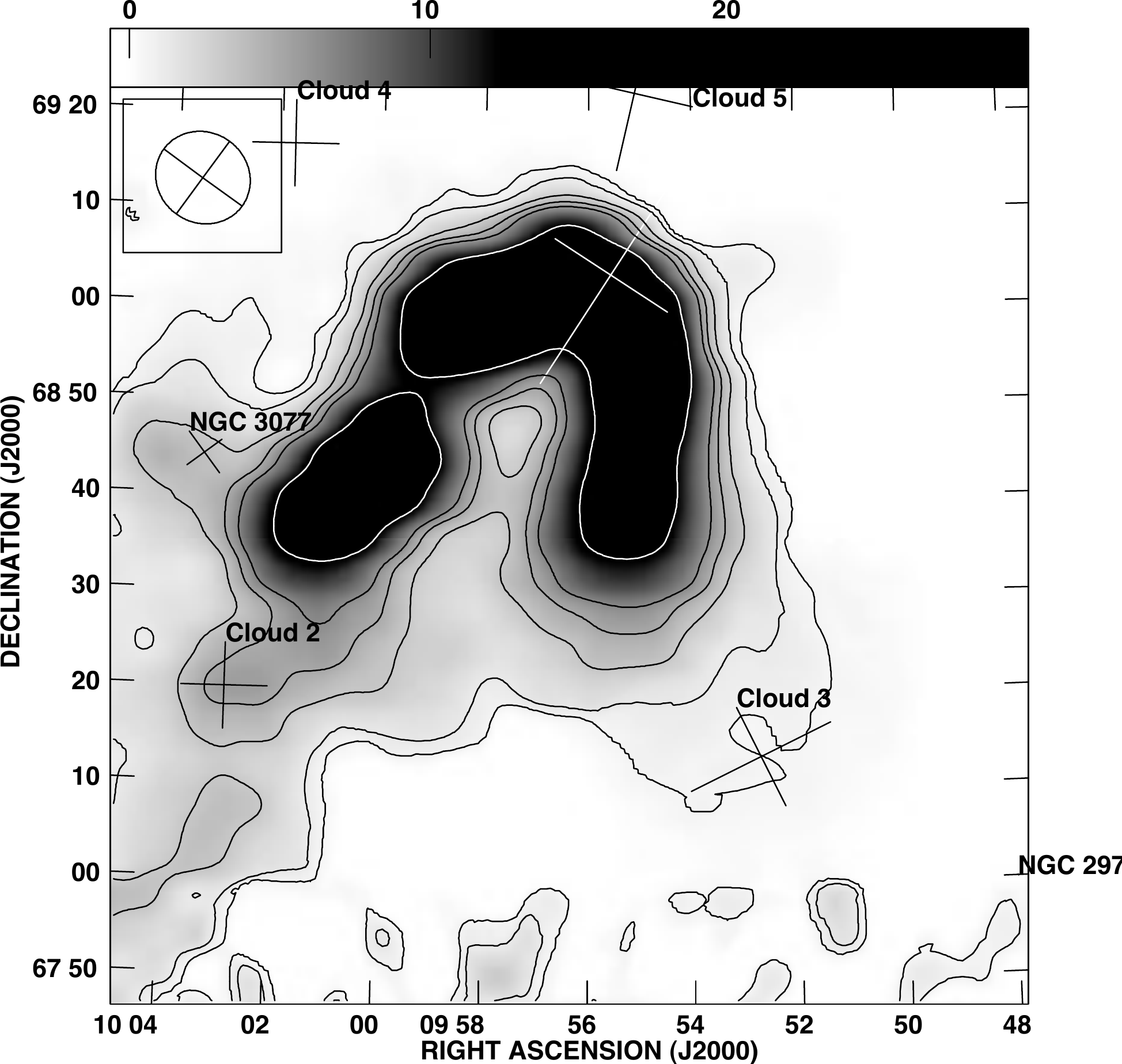} &
\includegraphics[width=0.333\linewidth, height=0.333\linewidth]{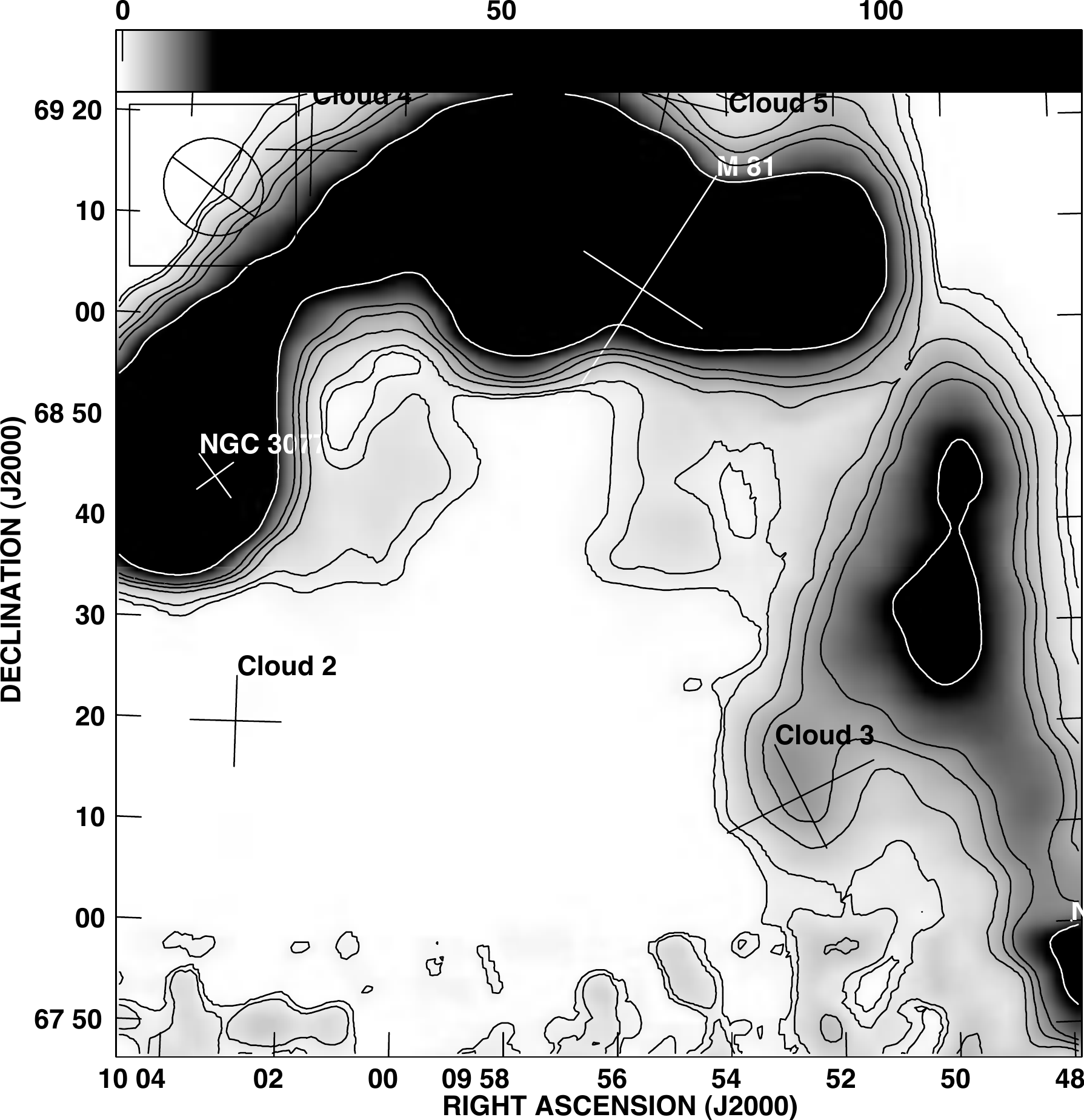}\\
Cloud 2 (-115 to -80 km s$^{-1}$)& Cloud 3 (5 to 75 km s$^{-1}$) \\
\includegraphics[width=0.333\linewidth, height=0.333\linewidth]{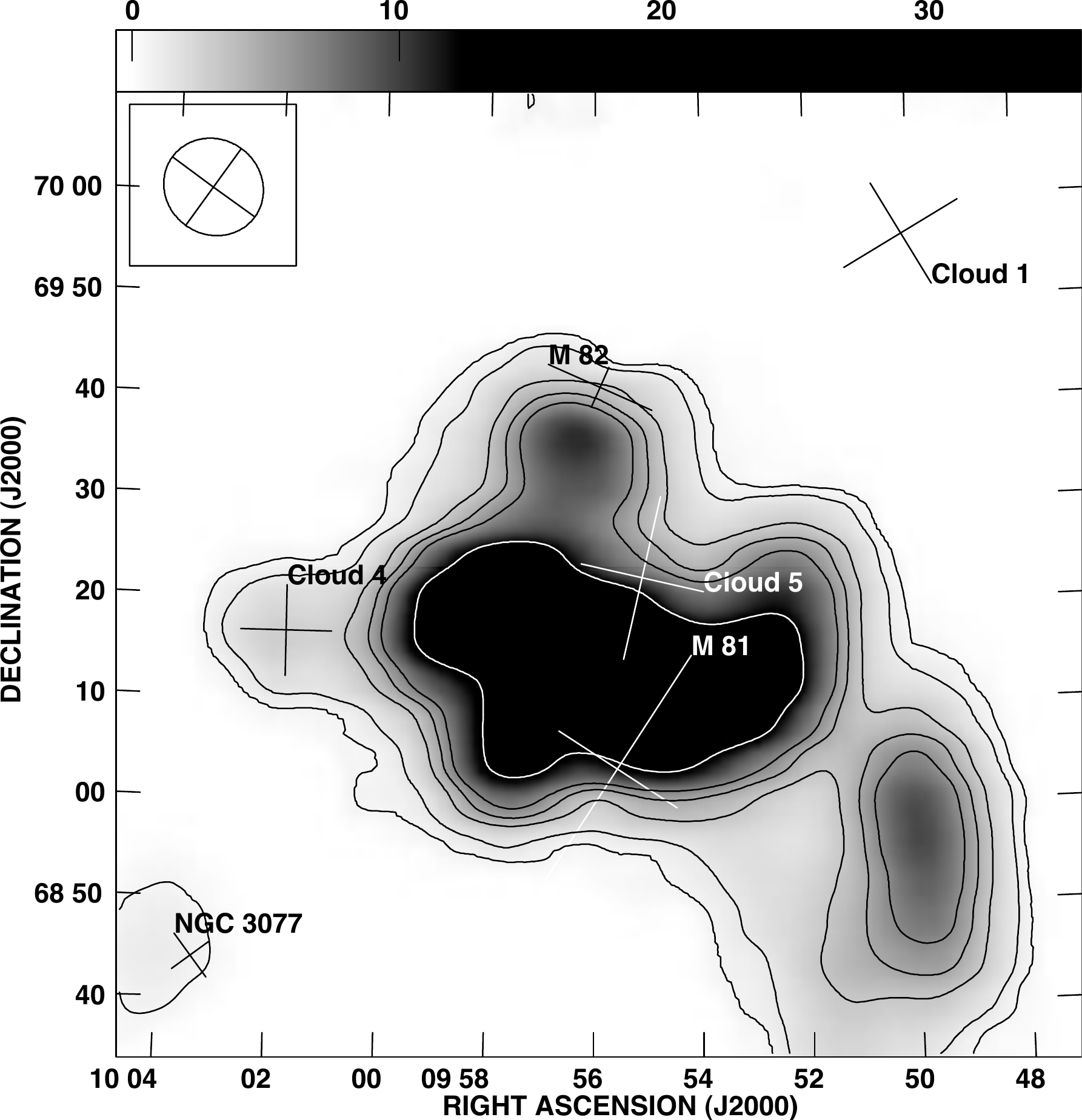}  & 
\includegraphics[width=0.333\linewidth, height=0.333\linewidth]{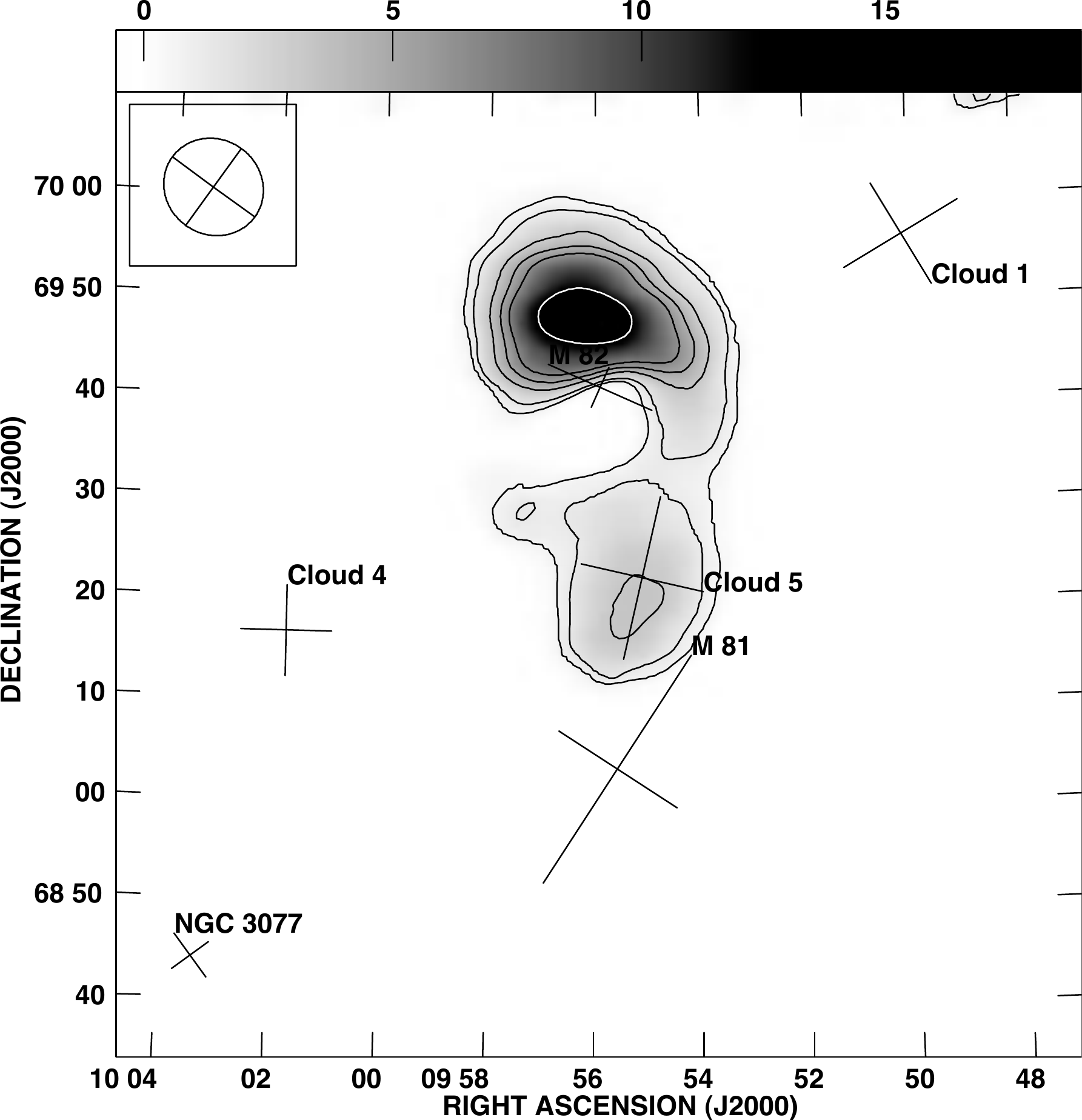} \\
Cloud 4 (75 to 95 km s$^{-1}$)& Cloud 5 (270 to 320 km s$^{-1}$)\\
\end{tabular}
\caption{Contour maps of each \hone cloud candidate, overlaid with \hone column density maps of the velocity range in which each cloud is detected. The greyscale is 0 to 12.5 kJy/beam $\times$ km/sec. Contour levels are 65 Jy/beam $\times$ km/sec $\times$ (10, 20, 50, 75, 100, 200, 500, 10000). The top left-hand panel shows the zoom regions for the contour plots; the upper box includes Clouds 1, 4, and 5 and the lower box includes Clouds 2 and 3. Below each map, the velocity range summed into the integrated intensity map is indicated.}
\label{contall3}
\end{figure}

The column density  was calculated for the entire field observed and each new \hone cloud.  Moment maps showing the column density are presented in Figures \ref{totalmoment} and \ref{contall3}.  For the observed field and major 
galaxies, the flux values at velocities from -250 to 340 km s$^{-1}$ were summed. For new objects, the flux in the velocity range over which each object detected was summed. The flux vs. velocity profile of the object was obtained from the AIPS task ISPEC. 

\subsection{Total Mass over the Field}

\begin{figure}
\epsscale{0.8}
\plotone{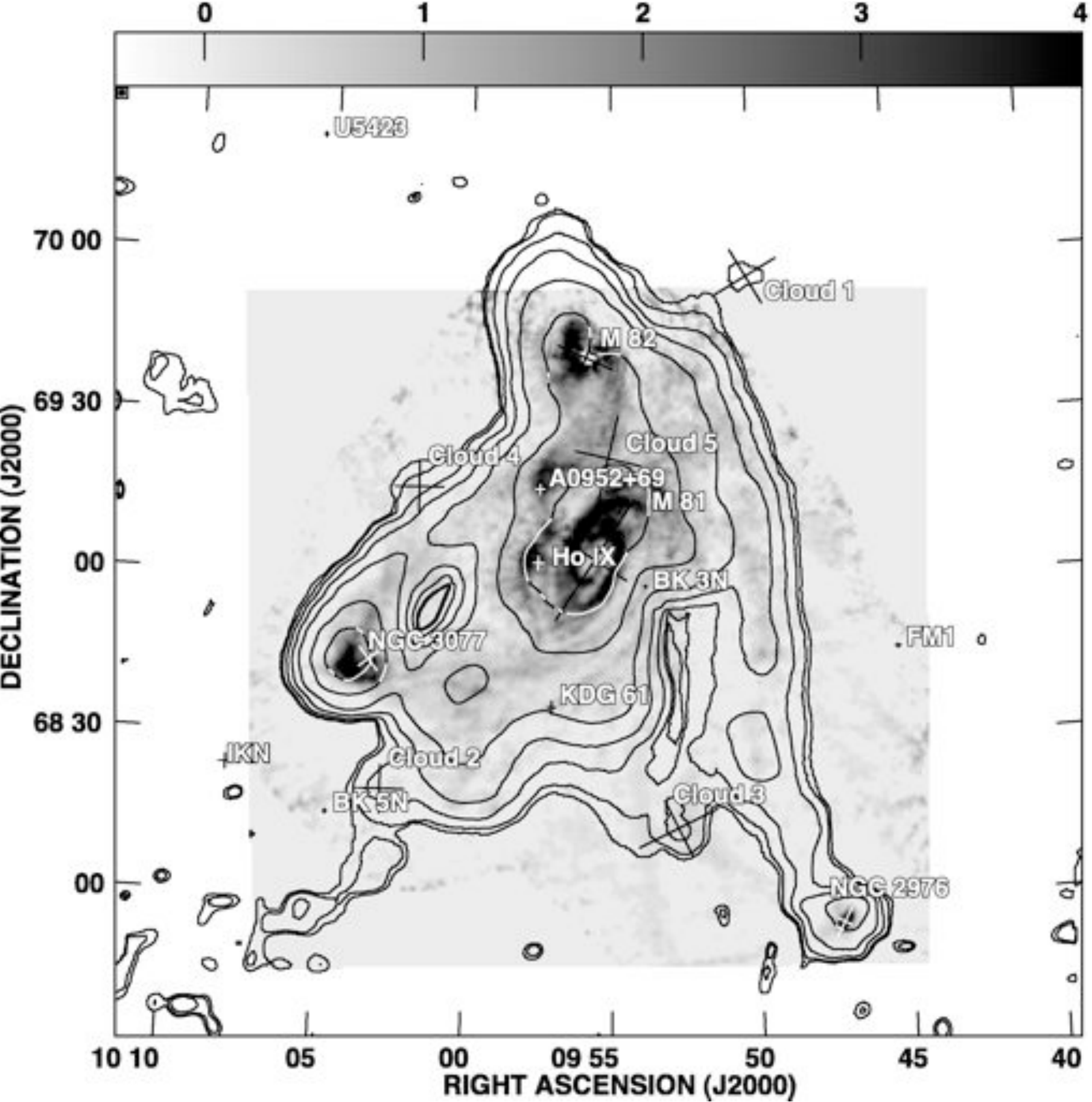}
\caption{The contours show the GBT \hone integrated column density map of the field observed. Contours are overlaid at 1500,  3000,  7500, 15000,  30000,  75000, 150000 and 300000 kJy/beam $\times$ km/sec. The velocity range included is -250 to 340 km s$^{-1}$. The integrated column density map from the VLA observations of  \citet{yun94} is shown in greyscale over the range  -400 to 4000 kJy/beam $\times$ km/sec. }
\label{totalmoment}
\end{figure}

Neutral hydrogen was detected over the velocity range -250 to 340 km s$^{-1}$, and these channels are summed and shown as contours in Figure \ref{totalmoment}. The integrated column density map from \citet{yun94} is shown as the grey scale image in Figure \ref{totalmoment}. The locations of each major galaxy, dwarf galaxy, and  \hone cloud detection are indicated. 

The spectrum of the entire field used to compute the neutral hydrogen mass of the group is shown in Figure \ref{glx_ispec}. The computed mass was 10.46 $\pm$ 2.86 $\times$ 10$^{9}$ M$_{\odot}$. For this mass and all following mass estimates, the error estimates are 1$\sigma$.   The dominant error contribution is the uncertainty in the absolute calibration.  Our total mass value is greater than the values found by \citet{app81} and \citet{yun99} . Our study is more sensitive to faint, extended emission over the entire field than either of these previous studies.

\subsection{Major Galaxies}

\begin{figure}
\centering
\begin{tabular}{cc}
\includegraphics[width=2.75in]{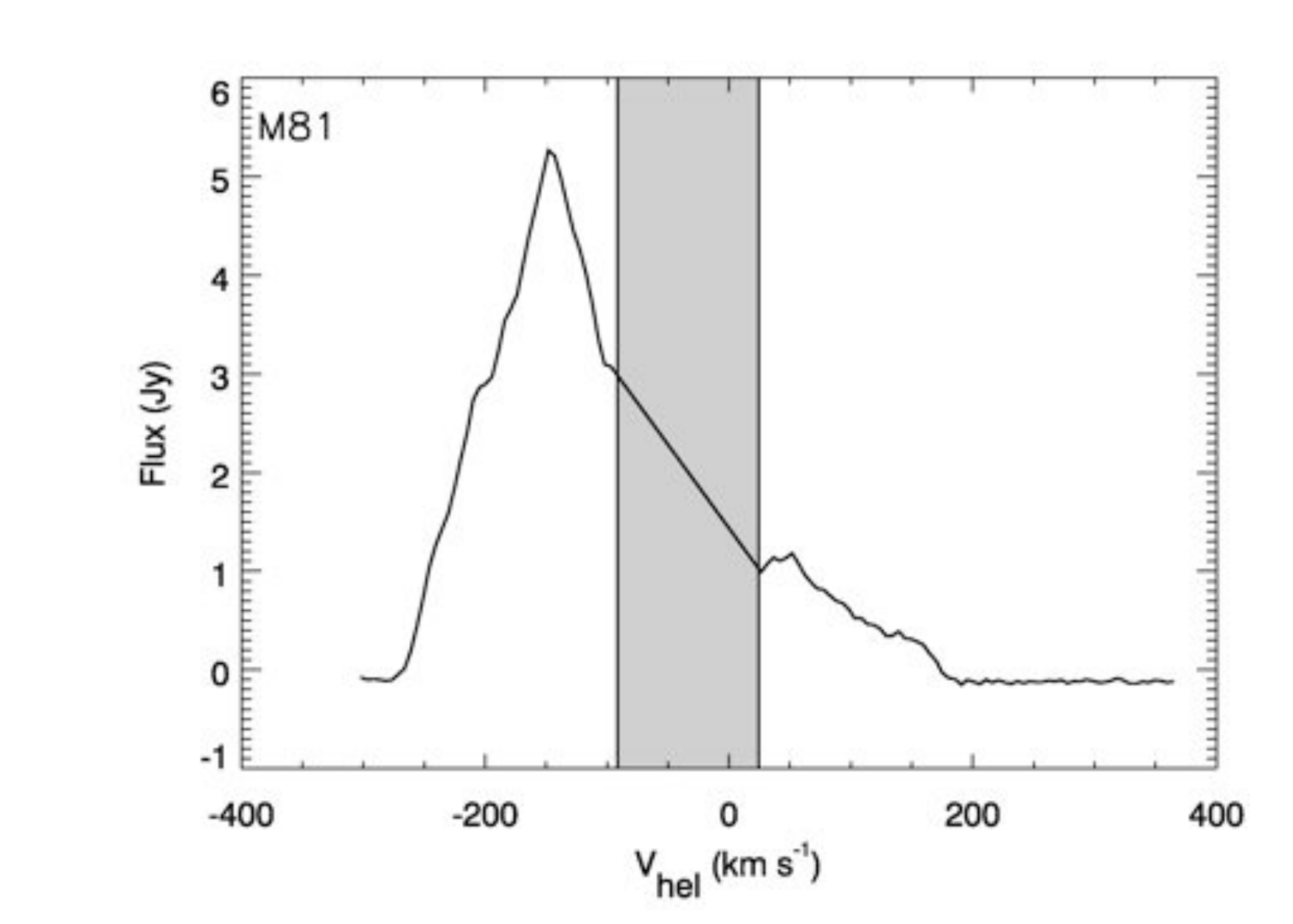}  &
\includegraphics[width=2.75in]{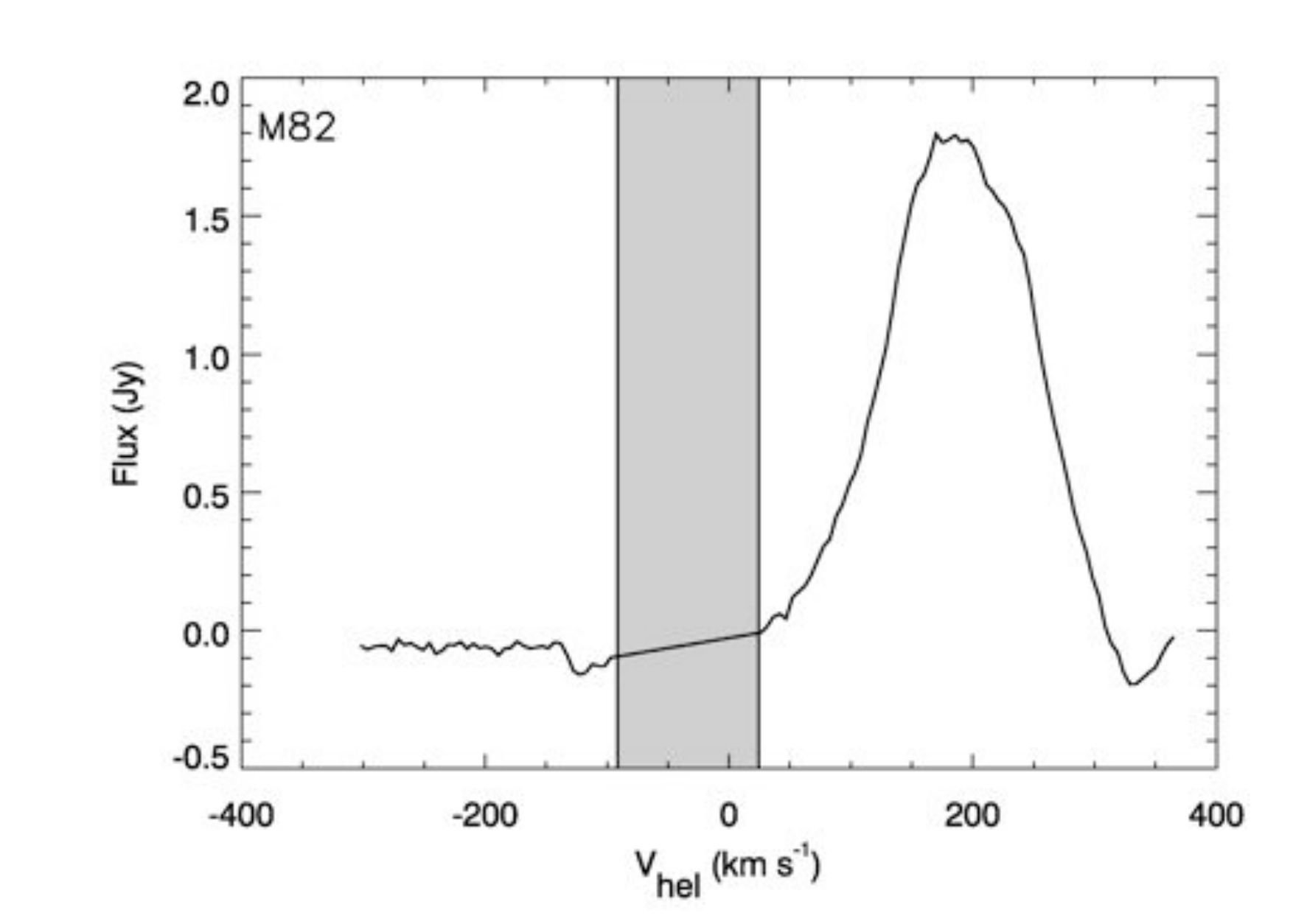}  \\
\includegraphics[width=2.75in]{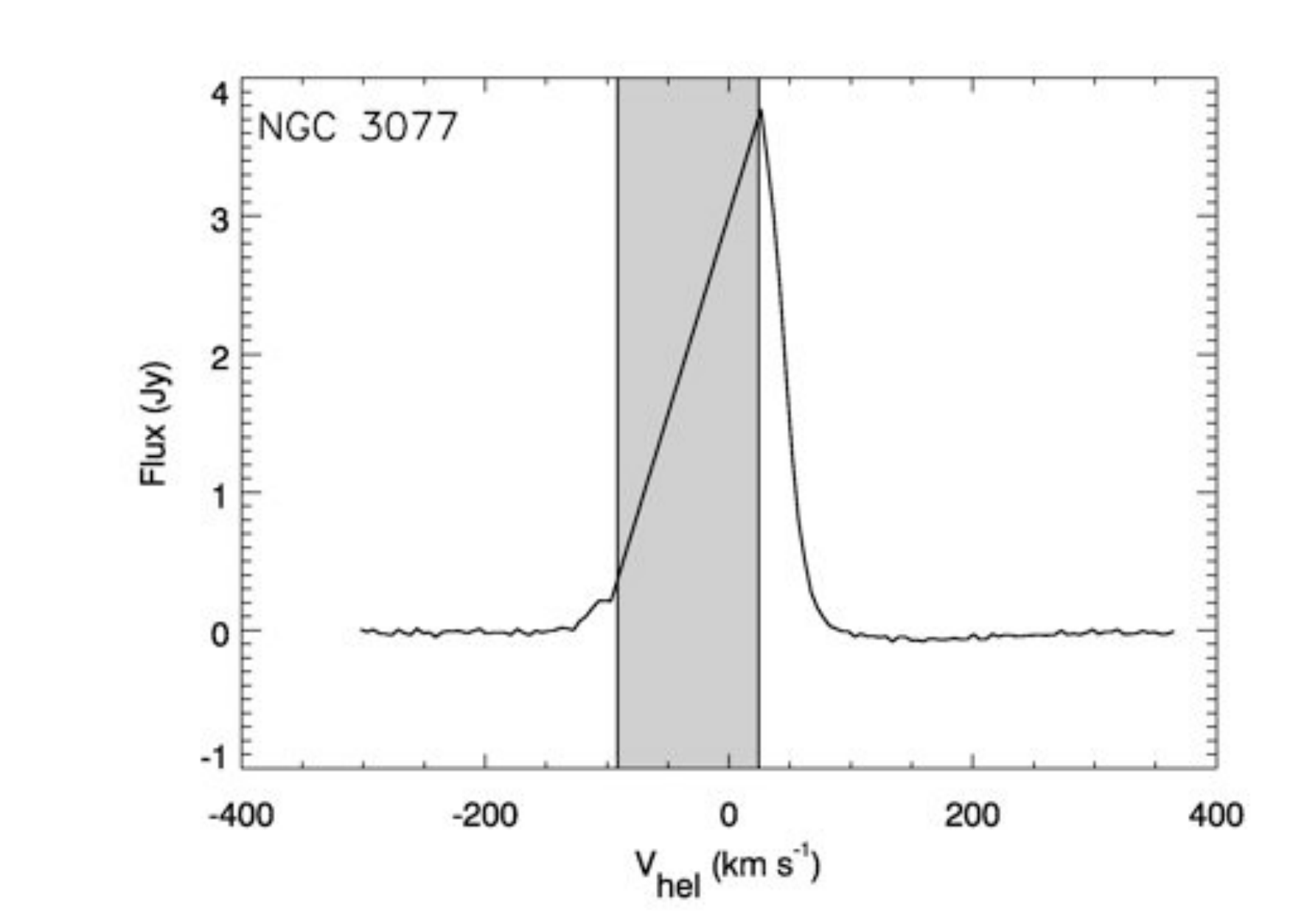}  &
\includegraphics[width=2.75in]{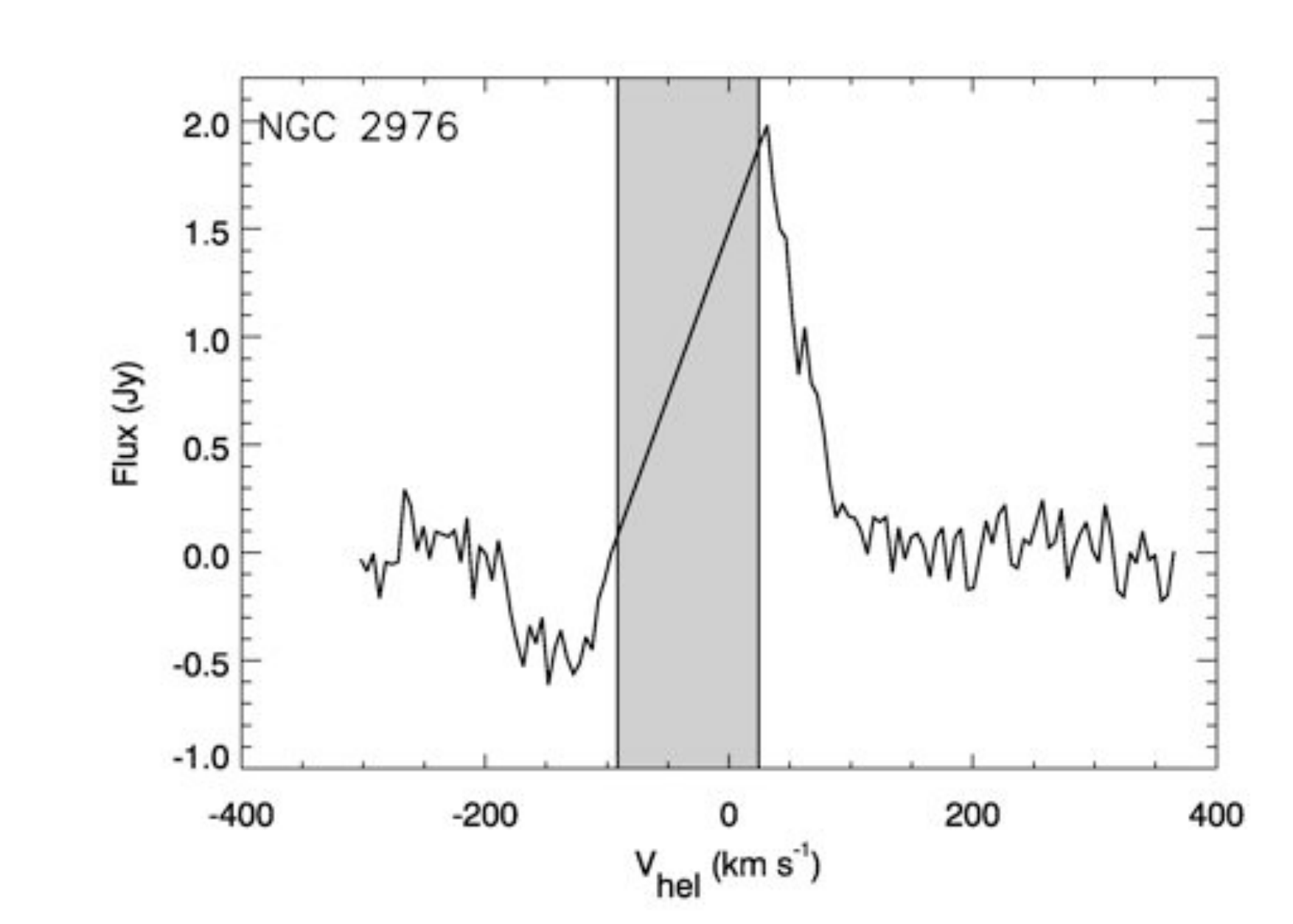}  \\
\includegraphics[width=2.75in]{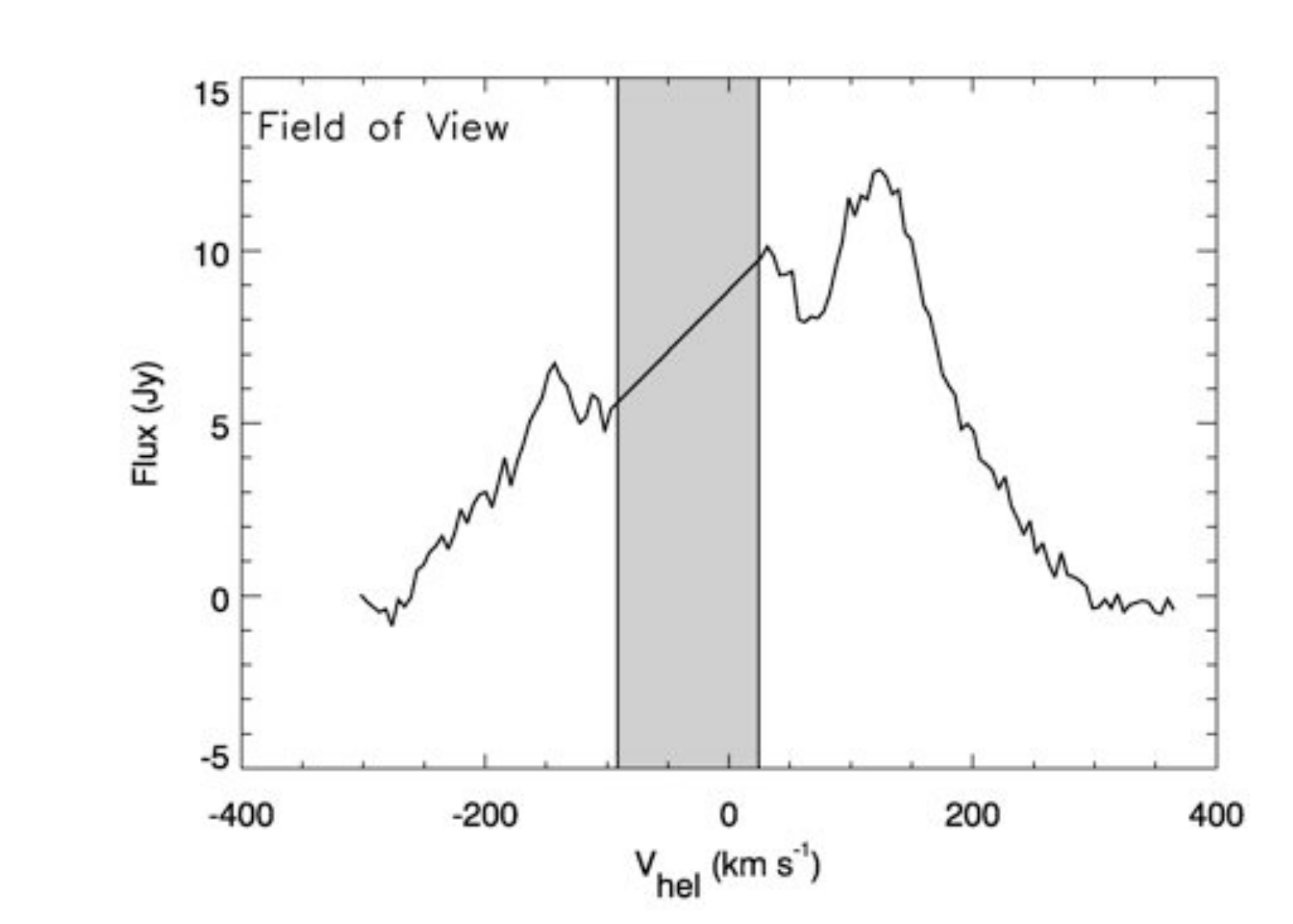} & \\
\end{tabular}
\caption{\hone intensity profiles of major group galaxies, used to calculate \hone mass. From top left: M81, M82, NGC 3077, and NGC2976. Also included is the \hone spectrum of the entire area observed (bottom left). Note that the velocity range from -85 to 25 km s$^{-1}$ is interpolated to remove foreground gas}
\label{glx_ispec}
\end{figure}

The \hone spectra for the major galaxies are presented in Figure \ref{glx_ispec}. Neutral hydrogen masses of known galaxies were calculated to confirm the accuracy of our mass calculation methods. Our calculated values for 
$M_{HI}$ are in agreement with the literature; see Table 4. The dwarf galaxy BK 1N  \citep{huc98} is also detected by our observations in the velocity range $\sim$ 560 to 580 km s$^{-1}$.  BK 1N has a velocity outside the range shown in Figure \ref{glx_ispec}.
 
\section{New Identifications} 
\label{ressec}
The newly detected clouds were identified by individual examination, by eye, of all spectral line images for this
region in the velocity range -605 to -85 km s$^{-1}$ and 25 to 1970 km s$^{-1}$. We detect five neutral hydrogen clouds associated with the M81 group of galaxies. See Tables 5 and 6 for a summary of detections. Three of the clouds are clearly associated with a particular galaxy, and two are located between two known galaxies.  All of the clouds have a velocity close to that of at least one of their parent galaxies. Clouds were only detected between -105 and 280 km s$^{-1}$, and the maximum difference between a cloud and associated galaxy was 119 km s$^{-1}$.

Figure \ref{contall3} shows \hone column density maps of each detection.  For these maps, only the velocity range in
which the cloud of interest was detected is summed into the integrated intensity map.  One channel on each end of the detected range was excluded from the integrated intensity map in order to increase the signal to noise ratio and highlight the cloud. Additionally, all of the clouds are within 35 kpc (in projection) of the group. However, this value should be taken as an estimate, as the exact distance to each galaxy in the group, and more importantly, the distance to each cloud, is uncertain. In any case, the approximate distance values are consistent with detected distances of clouds near M31 \citep{thilker04}. 

\begin{figure}
\centering
\begin{tabular}{cc}
\includegraphics[width=3in]{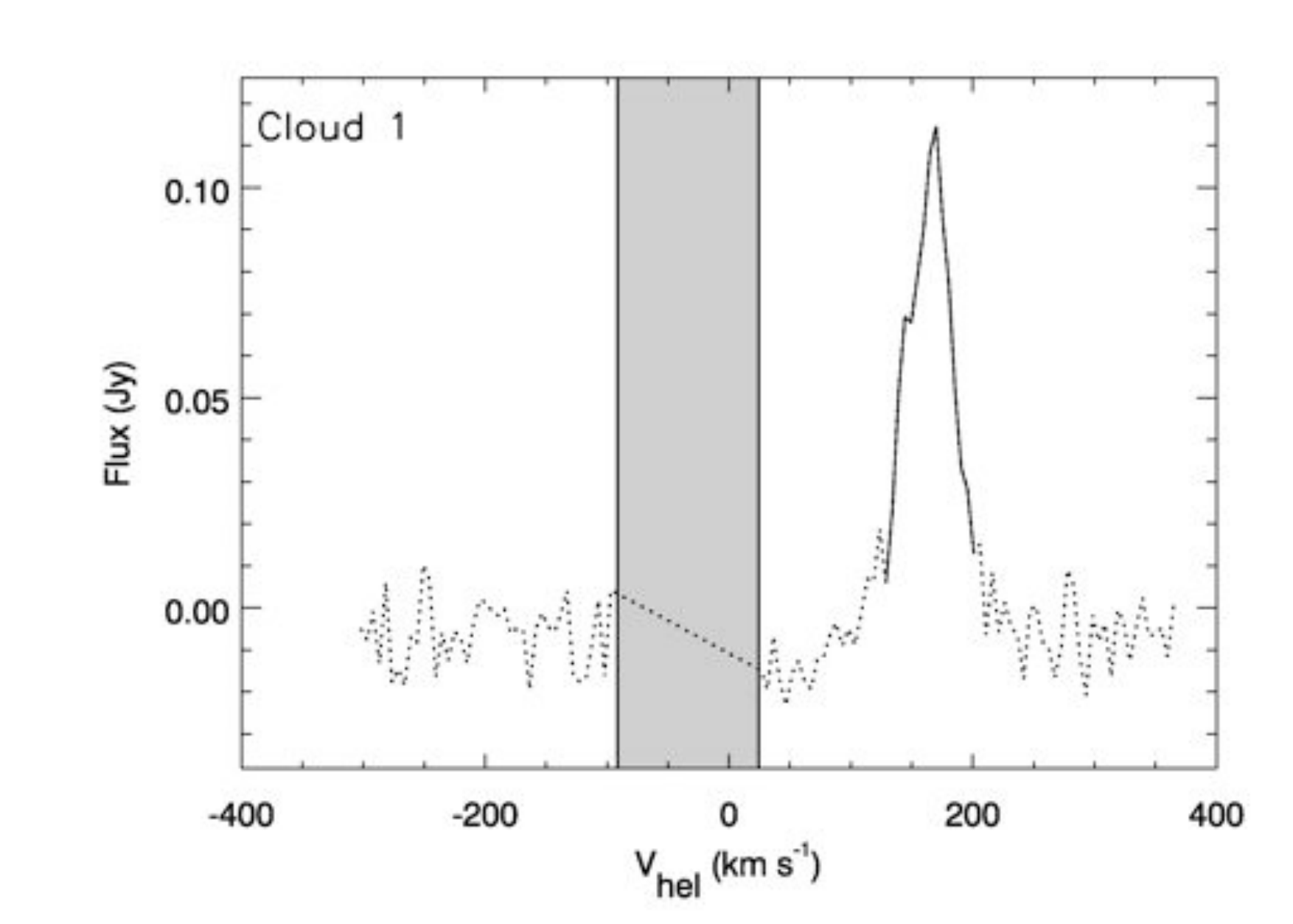}  &
\includegraphics[width=3in]{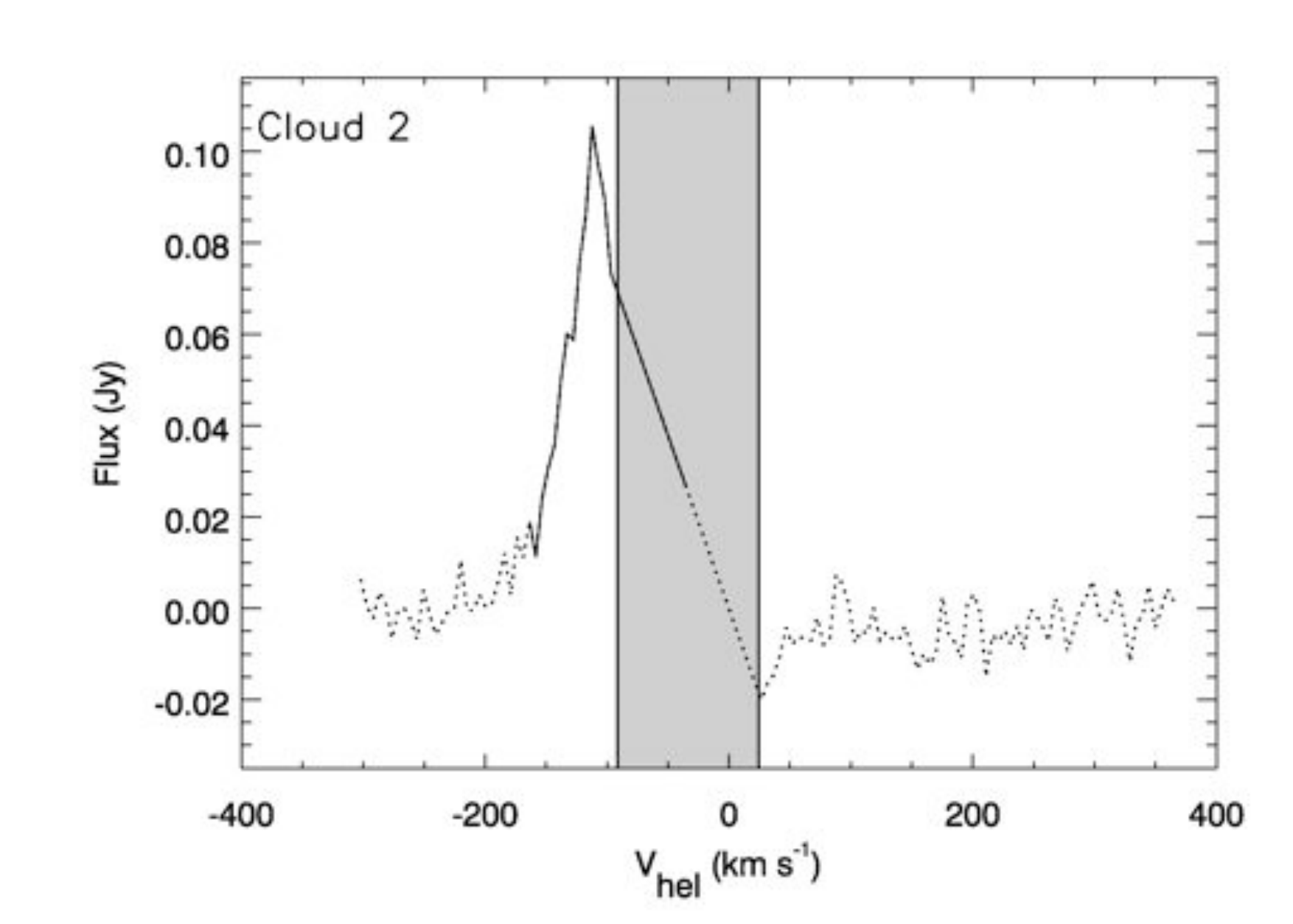} \\
\includegraphics[width=3in]{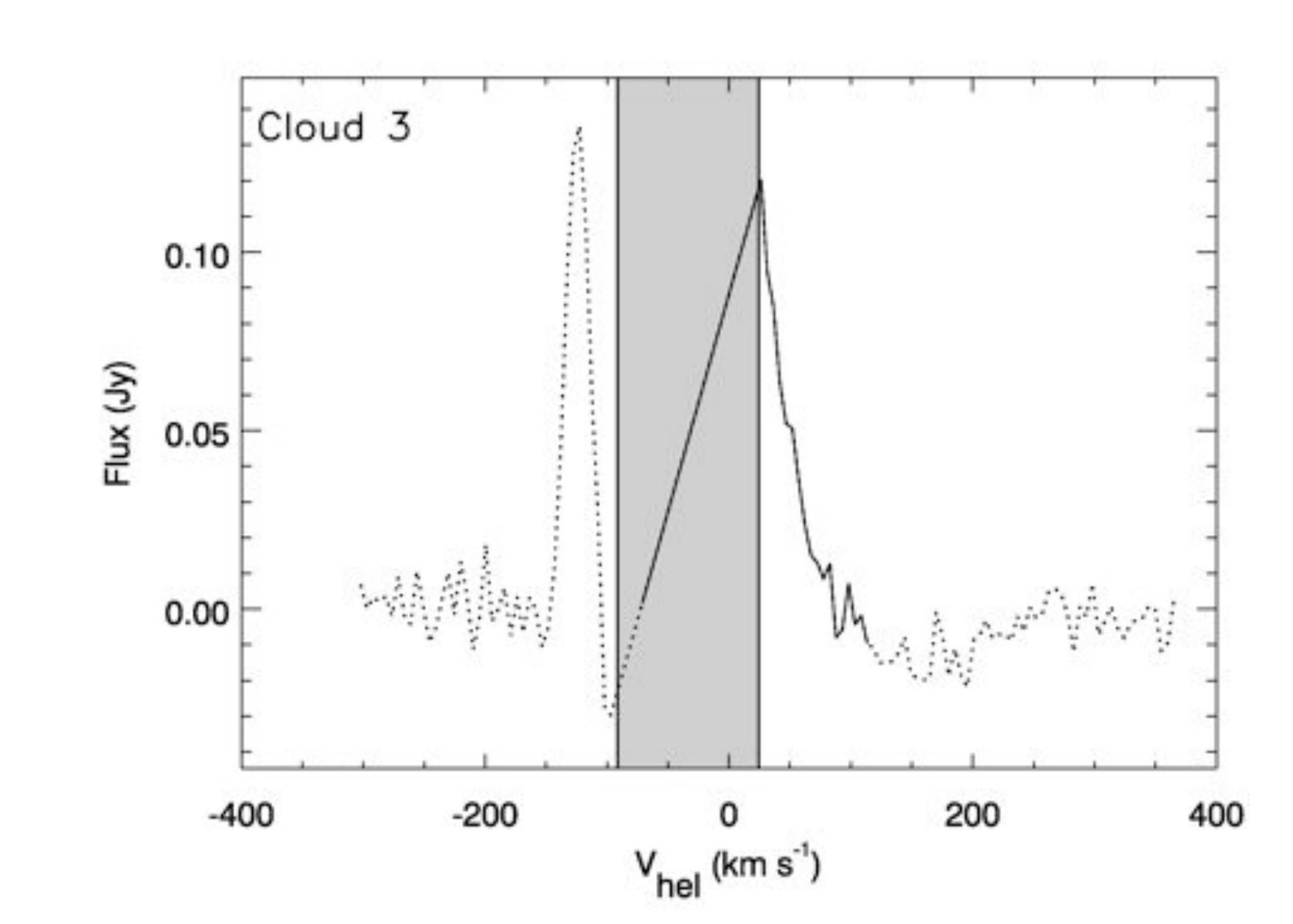}	&
\includegraphics[width=3in]{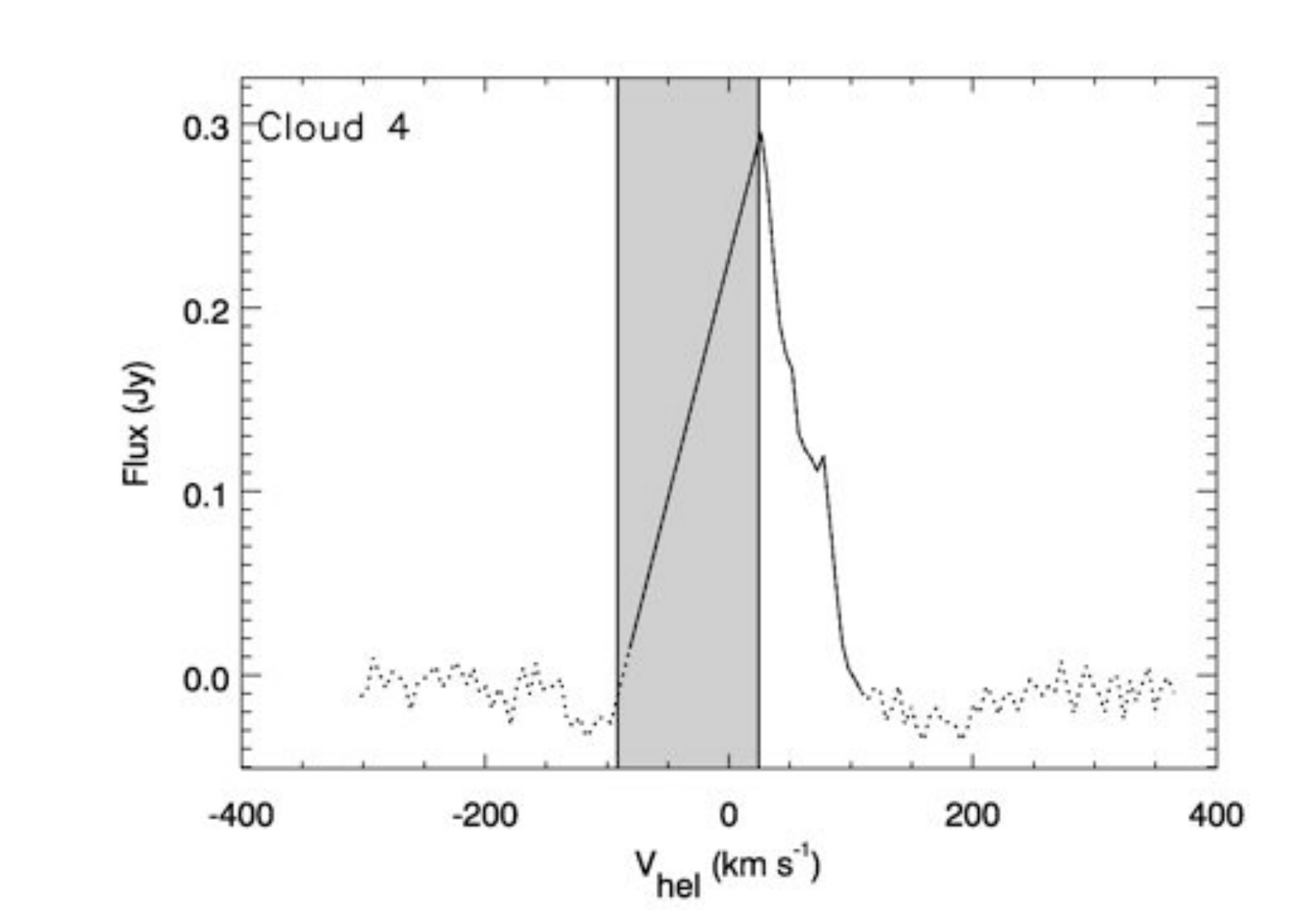} \\
\multicolumn{2}{l}{\includegraphics[width=3in]{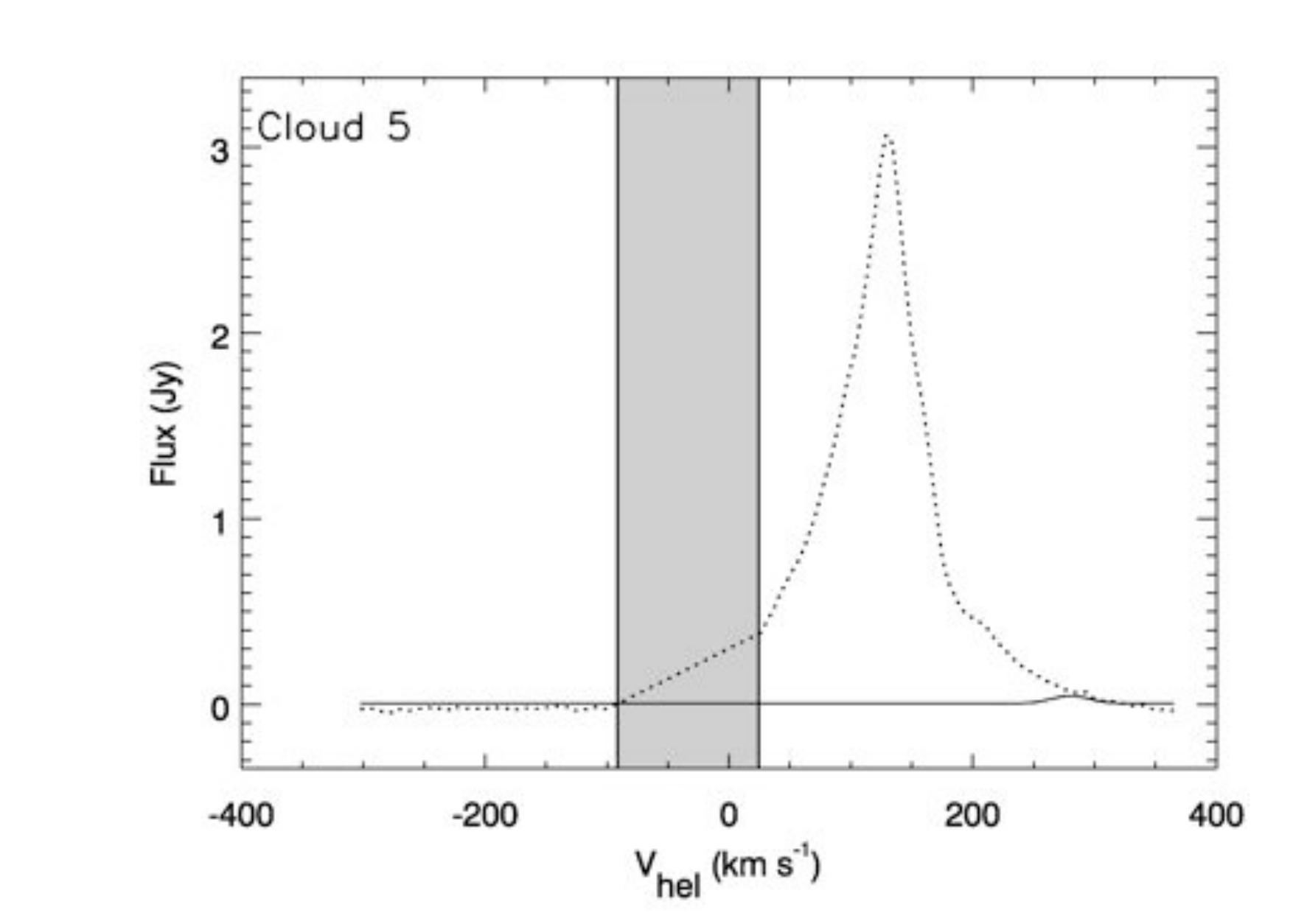}} \\
\end{tabular}
\caption{\hone profiles for all cloud detections. The solid line indicates the range that was summed in order to calculate the cloud mass. For Cloud 5, the data (dashed line) along with Gaussian fit for Cloud 5 (solid line) are plotted. Note that the velocity range from -85 to 25 km s$^{-1}$ is interpolated to remove foreground gas}
\label{ispec}
\end{figure}

Figure \ref{ispec} shows the spectra of each \hone cloud. The mass of each cloud was measured by summing the flux densities measure for the angular regions over the velocity range show in figure \ref{ispec}. The masses of the detected clouds are of order 10$^{7}$ $M_{\odot}$. In order to measure the systemic velocity and velocity dispersion of each cloud, a Gaussian function was fit to the average intensity profile. 

\begin{figure}
\centering
\begin{tabular}{cc}
\includegraphics[width=0.333\linewidth, height=0.333\linewidth]{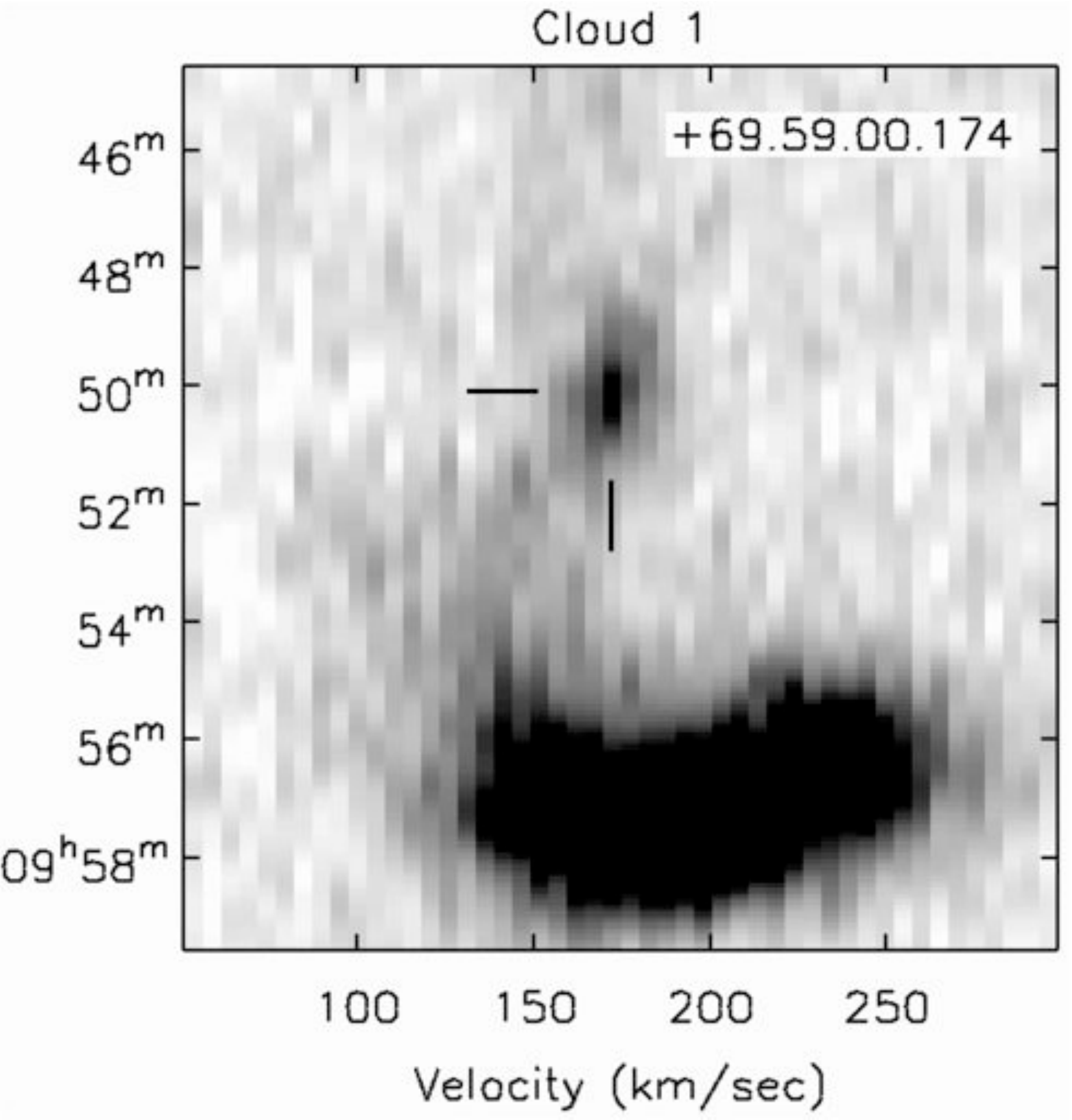}  &
\includegraphics[width=0.333\linewidth, height=0.333\linewidth]{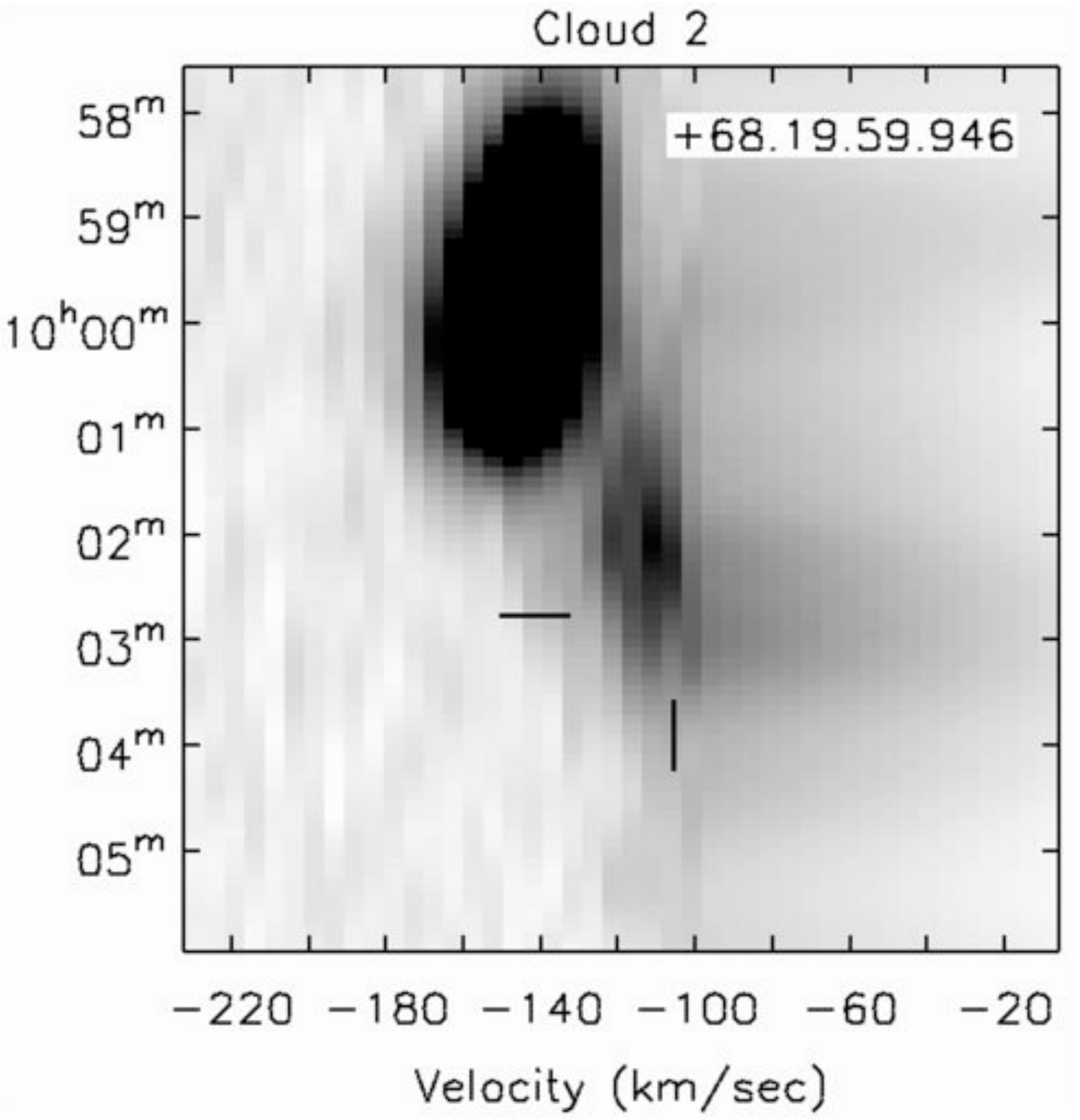} \\
\includegraphics[width=0.333\linewidth, height=0.333\linewidth]{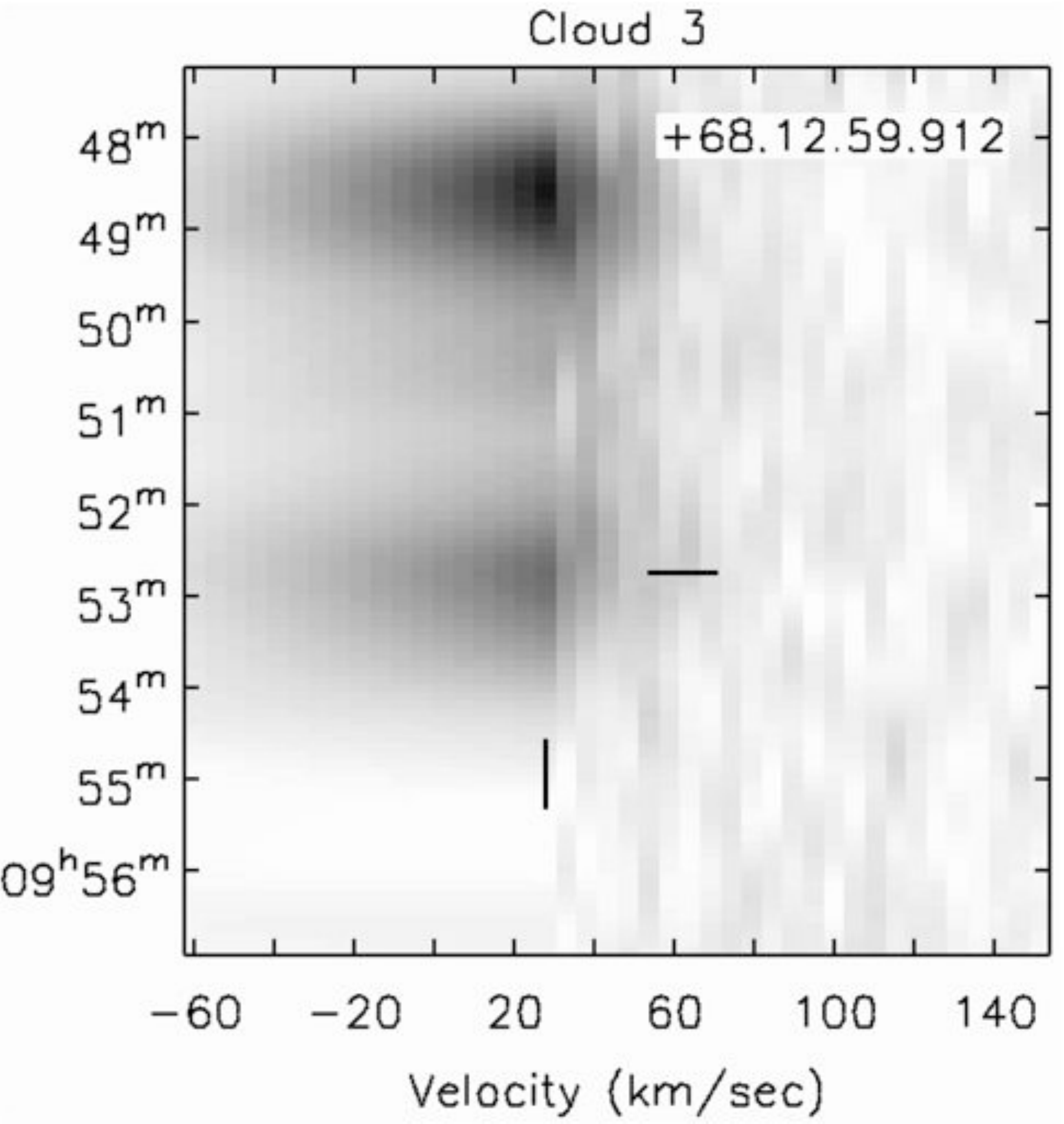}	&
\includegraphics[width=0.333\linewidth, height=0.333\linewidth]{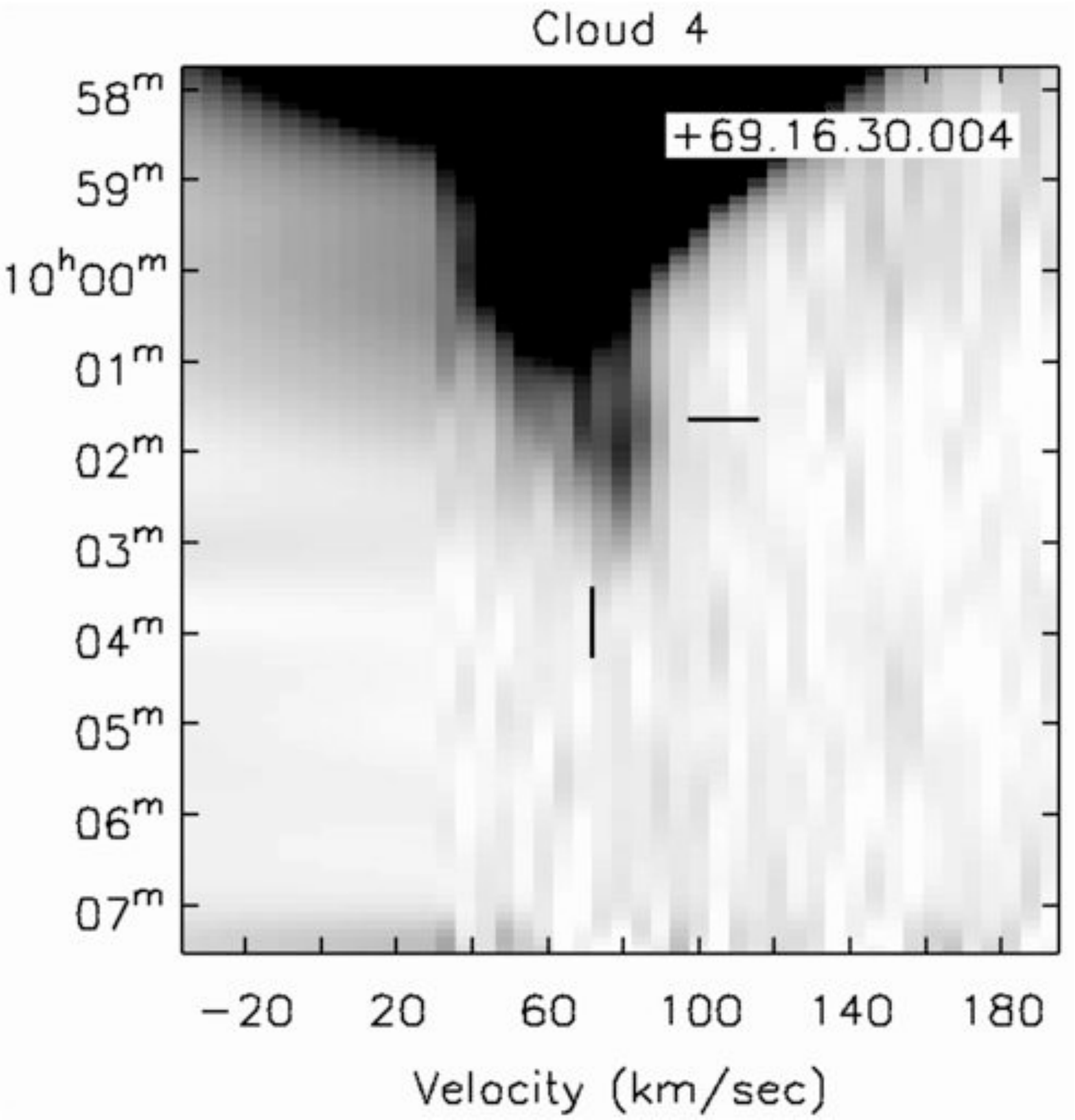} \\
\multicolumn{2}{l}{\includegraphics[width=0.333\linewidth,height=0.333\linewidth]{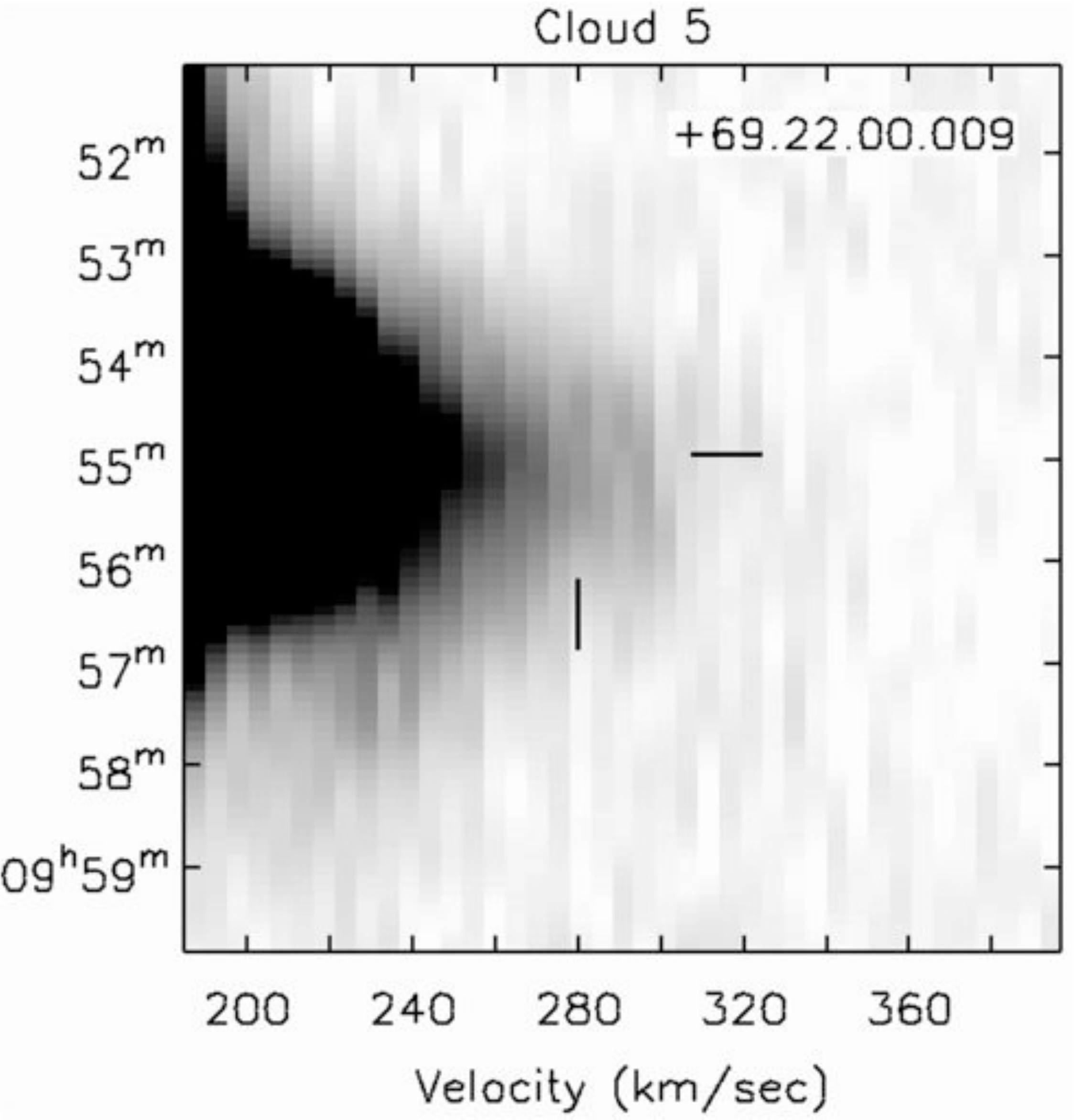}} \\
\end{tabular}
\caption{PV diagrams for all \hone cloud detections. Each cloud is located approximately in the center of the diagram, marked by pointers. Each diagram extends $\pm$ 100 km s$^{-1}$ from the cloud's peak velocity. The PV cut runs through the center of the clouds in declination, and the diagram is centered on the cloud in right ascension.  The greyscale runs from  -0.02 to 0.2 Jy beam$^{-1}$. The parent galaxy for each cloud are also visible in the PV diagrams.}
\label{pv}
\end{figure}

Figure \ref{pv} shows position-velocity (PV) diagrams for each new \hone cloud.  Whereas clouds may be difficult to distinguish from the extended \hone emission associated with major galaxies and tidal streams in the column density maps, the \hone spectra and position-velocity diagrams show the distinct nature of these new clouds.

We have compared the GBT observations with the VLA images from \citet{yun93}.   Below, we describe the \hone distribution found in the VLA images at the locations of each cloud. The combined GBT and VLA images are presented 
in \citet{langprep}, along with a description of the process of combining GBT and VLA images.

\subsection{Cloud 1}
Cloud 1 is located to the north-west of M82, separated by approximately 31 kpc in projection. It is clearly separated from M82. It is detected in the velocity range 132 to 204 km s$^{-1}$, with peak emission at 165 km s$^{-1}$. We calculate a mass of 1.47 $\pm$ 0.35 $\times$ 10$^{7}$ $M_{\odot}$ for Cloud 1. 

Cloud 1 lies outside the region imaged by \citet{yun93}, and no VLA images are currently available for this area.  
VLA observations are scheduled to supplement the previous work.

\subsection{Cloud 2}
Cloud 2 sits immediately to the south-east of NGC 3077, at a distance of 23 kpc. It is a distinct clump of \hone situated within a filamentary structure that extends from NGC3077 south and west, encompassing the dwarf galaxies BK5N, IKN, and K64. The cloud is in the upper range of the filament velocity. The morphology of the filament to the west of the cloud suggests that it may be plunging into NGC3077. Alternatively, the cloud may be gas blown out of the galaxy by star formation.

Cloud 2 is detected in the range -162 to -69 km s$^{-1}$, with peak emission at -105 km s$^{-1}$. We calculate a mass of 2.25 $\pm$ 0.49 $\times$ 10$^{7}$ $M_{\odot}$ for Cloud 2.

Cloud 2 is very weakly visible in the VLA images of \citet{yun93}.  Clearly the VLA images are missing diffuse emission associated with this cloud.  Other regions have similar strength in the VLA images, but are significantly weaker in the GBT images.   A weak secondary peak in the GBT image, at 10$^h$02$^m$12.3$^s$, 68$^{\circ}$20$^`$34$^{``}$, is also visible in the VLA image.

This cloud and the associated filament may be associated with the clouds in the same vicinity detected by \citet{brinks07} using the VLA. We will seek to confirm the detections by \citet{brinks07} with our scheduled VLA observations of the region.

\subsection{Cloud 3}
Cloud 3 is located 27 kpc to the north-east of NGC 2976. It is distinct from NGC 2976. Cloud 3 overlaps with the velocity range of the Galaxy, making properties difficult to determine accurately.  Cloud 3 is detected in the range -17 to 86 km s$^{-1}$, with peak emission at 11 km s$^{-1}$. We calculate a mass of  2.67 $\pm$ 0.65 $\times$ 10$^{7}$ $M_{\odot}$ for Cloud 3.

Cloud 3 is clearly resolved in the GBT image and only barely detected above the background noise in the VLA images of \citet{yun93}.  

\subsection{Cloud 4}
Cloud 4 is located to the west of the \hone bridge between M81 and M82. It is clearly bound to the \hone bridge, but is a distinct clump. Cloud 4 also overlaps with the velocity range of the Galaxy.  Cloud 4 is detected in the range -19 to 111 km s$^{-1}$, with peak emission at 72 km s$^{-1}$. We calculate a mass of 8.37 $\pm$ 1.75 $\times$ 10$^{7}$ $M_{\odot}$ for Cloud 4.

Although Cloud 4 is located equidistant from M81 and M82 at a projected distance of 33 kpc from both galaxies, it is most likely associated with M81, as it is much closer to M81 in velocity.  

In the \hone images of \citet{yun93}, cloud 4 has the most compact structure of our newly detected clouds.    The peak in the VLA emission in this region is very close to the center of the GBT detection.

\subsection{Cloud 5}
Cloud 5 is to the south of M82, located approximately halfway between M81 and M82 (separated by 18 kpc from both galaxies). 
Cloud 5 is only distinct from M82 in a small range of velocities.
It is unclear whether the cloud is actually a separate feature from M82; it appears as a red shoulder in the \hone profile. In order to calculated the mass of the cloud a Gaussian was fit to the \hone spectrum in the peak range where the cloud was detected. 
The cloud is detected from approximately 267 to 302 km s$^{-1}$, peaking at approximately 280 km s$^{-1}$. Since the \hone profile of the cloud is blended with M82, the calculated mass is difficult to determine. The results of the Gaussian fit give a mass of 0.69 $\pm$ 0.27 $\times$ 10$^{7}$ $M_{\odot}$.

Cloud 5 is a large diffuse region in the maps of \citet{yun93}, only a few sigma above the noise in the VLA images.

\subsection{Other Structures}
All \hone cloud detections have some level of \hone structure surrounding them, connecting them to the associated galaxy. In particular, Cloud 2 is a clump within a pronounced filament to the southwest of NGC 3077.  This cloud/filament structure may be associated with the dwarf galaxies BK5N, IKN, and K64, as they are close in projection.  The velocity of K64 is known, at ($v_{K64}$ = -18 km s$^{-1}$), however, the detected velocities of the cloud and filament are within about 90 km s$^{-1}$ of $v_{K64}$, so association is certainly possible.  

Cloud 4 appears to be quite tightly bound to the \hone bridge connecting M81 and M82. 

\section{Comparison With Numerical Simulation}
\label{simsec}

A useful insight on whether the newly discovered \hone clouds are part of the system of tidal debris produced by the recent collisions involving M81, M82, and NGC~3077 can be obtained by comparing their observed locations and kinematics with the results of a numerical simulation of the tidal interactions in this group.  Such a comparison can be made with the simulation of \citet{yun99}.

The simulation by \citet{yun99} is primarily an illustrative model in that only the velocity field of tidal debris is mapped using a limited three-body calculation, rather than employing a full N-body calculation.  On the other hand, this approach has been shown to be effective in tracing the location and kinematics of tidally disrupted \hone disks as the \hone emitting clouds originate from a dynamically cold structure \citep{tt72}.  This simulation suggests that nearly all
of the observed \hone features in the M81 group are consistent with being part of the tidally driven structures resulting from the collisions involving all three galaxies during the last 200-300 Myrs -- see \citet{yun99} for more detailed discussions.  Specifically, the locations of the Clouds 1, 2, 4, and 5 are coincident with the locations of tidally sprayed material from the individual disks.  While a chance coincidence cannot be ruled out, this comparison suggests that the origin of these \hone clouds are tidally disrupted disk material, rather than a kinematically distinct satellite in the group.  A more sophisticated, self-consistent future numerical simulation will offer an improved understanding on the nature of these \hone clouds.  

\section{Conclusions}
\label{consec}

We have detected 5 new \hone clouds in the M81/M82 galaxy group, all with associated filamentary structures of \hone gas. We do not find a population of clouds at velocities between -605 and -105 km s$^{-1}$ or between 280 and 1970 km s$^{-1}$. All of the objects have properties similar to those of clouds previously found in our galaxy and other nearby galaxies.  The newly detected clouds have small distance offsets from larger neighboring galaxies and also
have small velocity offsets. 

We have also measured the mass of group galaxies, obtaining values in agreement with previous measurements, and calculated masses for our new objects. The detected \hone clouds have masses ranging from 0.69 to 8.37 $\times$ 10$^{7}$ M$_{\odot}$.  These cloud masses are larger than the recently discovered clouds near M31, and further high angular resolution images with greater sensitivity will likely yield detection of lower mass clouds.   

The region was searched for clouds from -605 to -85 km s$^{-1}$ and 25 to 1970 km s$^{-1}$. With a 7$\sigma$ detection threshold of $9.6 \times 10^{5} M_{\odot}$, clouds were found only within 120 km s$^{-1}$ and 35 kpc in projection of the galaxy with which they are associated. Therefore, it is likely that these clouds are part of the current interactions ongoing in the galaxy group. 

Previous studies of \hone clouds have been focused on the Milky Way and M31. Both of these galaxies are in a fairly relaxed state, and not currently undergoing strong interaction with other galaxies. The M81/M82 group, in contrast, is obviously interacting. Therefore, the study of \hone clouds and other structures in the group is very important in order to understand the full lifetime of these structures in relation to galaxy interactions. Our results are inconsistent with
models of primordial \hone clouds falling into the cluster. Perhaps this conclusion may be extended to the HVCs in the Milky Way; however, the clouds must be studied further to make this comparison.

Numerical modeling of ongoing interactions in the group are in progress \citep{chyprep}. Simulations will improve on the model of \citet{yun99} by utilizing a fully self-consistent N-body model of the galaxy group, with particular emphasis on reproducing the isolated \hone structures observed.  Follow-up VLA observations our newly detected clouds are scheduled, and we are preparing a description of combined VLA and GBT images of these clouds. These higher angular resolution interferometer images will be combined with GBT data to facilitate numerical modeling. Observations at higher spatial resolution will also be useful in order to probe the fine structure of these objects and detect objects with a smaller angular size. 

\acknowledgments

{\it Facilities:} \facility{GBT}.

\onecolumn

\begin{table}
\begin{center}
\caption{Known Galaxies in the M81 Group\tablenotemark{a}}
\begin{tabular}{l|c|c|c|c|c|c|c}
\tableline\tableline
Name 	& $\alpha$\tablenotemark{b}	& $\delta$\tablenotemark{b}	& V$_{hel}$\tablenotemark{c}	& Type\tablenotemark{d}	& D$_{maj}$\tablenotemark{b}	& D$_{min}$\tablenotemark{b}	& Inclination\tablenotemark{e} \\ \hline
	& (J2000)			& (J2000)			& (km s$^{-1}$)			&			& `				& `				& $^{\circ}$ E of N 	\\
\textbf{M 81} 			& 09$^h$55$^m$33.5$^s$	& 69$^{\circ}$03$^`$06$^{``}$     		& -34   & Sb    & 26.9  & 14.0  & 147 \\
Ho IX 				& 09$^h$57$^m$32.4$^s$	& 69$^{\circ}$02$^`$35$^{``}$     		& 46    & Irr   & 2.5   & 2.0   & ...   \\
BK 3N 				& 09$^h$53$^m$48.5$^s$	& 68$^{\circ}$58$^`$09$^{``}$     		& -40   & Irr   & 0.5   & 0.4   & ...   \\
A0952+69 			& 09$^h$57$^m$29.0$^s$	& 69$^{\circ}$16$^`$20$^{``}$     		& 100   & Irr   & 1.8   & 1.6   & ...   \\
KDG 61 				& 09$^h$57$^m$02.7$^s$	& 68$^{\circ}$35$^`$30$^{``}$     		& -116  & Sph   & 2.4   & 1.4   & ...   \\
\textbf{M 82} 			& 09$^h$55$^m$53.9$^s$	& 69$^{\circ}$40$^`$57$^{``}$     		& 203   & Irr   & 11.2  & 4.3   & -114  \\
\textbf{NGC 3077}		& 10$^h$03$^m$21.0$^s$	& 68$^{\circ}$44$^`$02$^{``}$     		& 14    & Irr   & 5.4   & 4.5   & -142  \\ 
Garland\tablenotemark{d} 	& 10$^h$03$^m$42.7$^s$ 	& 68$^{\circ}$41$^`$27$^{``}$     		& 50    & Im    & ...   & ...   & ...   \\	
BK 1N\tablenotemark{f}		& 09$^h$45$^m$14.3$^s$	& 69$^{\circ}$23$^`$23$^{``}$		& 571	& S	& ...	& ...	& ... \\
FM 1 				& 09$^h$45$^m$10.0$^s$ 	& 68$^{\circ}$45$^`$54$^{``}$    		 & ...   & Sph   & 0.9   & 0.8   & ...   \\
BK 5N 				& 10$^h$04$^m$40.3$^s$ 	& 68$^{\circ}$15$^`$20$^{``}$     		& ...   & Sph   & 0.8   & 0.6   & ...   \\
IKN\tablenotemark{g} 		& 10$^h$08$^m$05.9$^s$ 	& 68$^{\circ}$23$^`$57$^{``}$     		& ...   & Sph   & 2.7   & 2.3   & ...   \\	
\textbf{NGC 2976}\tablenotemark{g} & 09$^h$47$^m$15.6$^s$ & 67$^{\circ}$54$^`$49$^{``}$     	& 3     & Sm    & 5.9   & 2.7   & -119  \\
U5423\tablenotemark{g} 		& 10$^h$05$^m$30.6$^s$ 	& 70$^{\circ}$21$^`$52$^{``}$    	& 348   & BCG   & 0.9   & 0.6   & ...   \\	
KDG 64\tablenotemark{g} 	& 10$^h$07$^m$01.9$^s$ 	& 67$^{\circ}$49$^`$39$^{``}$     		& ...   & Sph   & 1.9   & 0.9   & ...   \\
KK 77				& 09$^h$50$^m$10.0$^s$	& 67$^{\circ}$30$^`$24$^{``}$     	& ...   & Sph   & 2.4   & 1.8   & ... \\ 		
IC 2574\tablenotemark{g}	& 10$^h$28$^m$22.4$^s$	& 68$^{\circ}$24$^`$58$^{``}$    	& 57    & Sm    & 13.2  & 5.4   & 50  \\
\tableline
\tablenotetext{a}{Within observed area}
\tablenotetext{b}{From \citet{kar04}}
\tablenotetext{c}{From \citet{yun99} for major galaxies, \citet{kar04} for others}
\tablenotetext{d}{From \citet{kar02}}
\tablenotetext{e}{From \citet{app81}}
\tablenotetext{f}{From \citet{huc98}.}
\tablenotetext{g}{May not be contained entirely within our search area}
\end{tabular}
\end{center}
\end{table}

\begin{table}
\begin{center}
\caption{Parameters of the Robert C. Byrd Green Bank Telescope System}
\begin{tabular}{ll}
\tableline\tableline
Telescope: \\
\hskip 3mm Diameter........................... & 100 m \\
\hskip 3mm Beamwidth (FWHM)................... &  9.1 \arcmin \\
\hskip 3mm Linear resolution ................. &  2.7 $D_{Mpc}$ kpc \\
\hskip 3mm Aperture efficiency  .............. &  $\sim 0.69$ \\
Receiver: \\
\hskip 3mm System Temperature ................ & $\sim 19.5$ K \\
Spectrometer: \\
\hskip 3mm Bandwidth.......................... & 50 MHz (10550 km s$^{-1}$) \\
\hskip 3mm Resolution, Hanning Smoothed....... & 6.1 kHz (1.29 km s$^{-1}$) \\
\tableline
\end{tabular}
\end{center}
\end{table}

\begin{table}
\begin{center}
\caption{Observations Summary}
\begin{tabular}{ll}
\tableline\tableline
Area: \\
\hskip 3mm $\alpha$ range (J2000).................. & 08$^h$25$^m$31.9$^s$ - 11$^{\circ}$25$^`$31.9$^{``}$ \\
\hskip 3mm $\delta$ range (J2000).................. & 67$^h$33$^m$25.7$^s$ - 70$^{\circ}$33$^`$25.7$^{``}$ \\
Observations: \\
\hskip 3mm Center Frequency (MHz).................. & 1410 \\
\hskip 3mm Bandwidth (MHz)......................... & 50 \\
\hskip 3mm Channel Width (kHz)..................... & 24.4 \\
\hskip 3mm Velocity Resolution (km s$^{-1}$)....... & 5.2 \\
Integration time (hours): \\
\hskip 3mm Inner Region............................ & 3.88 \\
\hskip 3mm Expanded Area........................... & 6.75 \\
Typical RMS noise (mK/channel): \\
\hskip 3mm Inner Region............................ & 18 \\ 
\hskip 3mm Expanded Area........................... & 51 \\
Sensitivity to \hone (1$\sigma$): \\
\hskip 3mm N$_{H}$ ($\times$ 10$^{17}$) \\
\hskip 6mm Inner Region............................ & 2.5 \\
\hskip 6mm Expanded Area........................... & 7.0  \\
\hskip 3mm M$_{H}$ ($\times$ 10$^{5}$ M$_{\odot}$) \\
\hskip 6mm Inner Region............................ & 1.4  \\
\hskip 6mm Expanded Area........................... & 3.9  \\        
\end{tabular}
\end{center}
\end{table}

\begin{table}
\begin{center}
\caption{Calculated Galaxy Masses}
\begin{tabular}{l|c|c|c}
\tableline\tableline
Galaxy	& Mass				& Mass from \citet{yun99}	& Mass from \citet{app81} \\
	& ($\times 10^9 M_{\odot}$)	&	 			& \\
\tableline
Field of Study & 10.46 $\pm$ 2.86        & 5.6                           & 5.4 \\
M81	& 2.67 $\pm$ 0.55		& 2.81 $\pm$ 0.56		& 2.19 $\pm$ 0.22 \\
M82	& 0.75 $\pm$ 0.16		& 0.80 $\pm$ 0.16		& 0.72 $\pm$ 0.07 \\
NGC3077	& 1.01 $\pm$ 0.21		& 0.69 $\pm$ 0.14		& 1.00 $\pm$ 0.10 \\
NGC2976 & 0.52 $\pm$ 0.11		& ...				& 0.16 $\pm$ 0.02 \\
\tableline
\end{tabular}
\end{center}
\end{table}

\begin{table}
\begin{center}
\caption{\hone Cloud Detections: Observed Properties}
\begin{tabular}{l|c|c|c|c}
\tableline\tableline
Cloud	& Association\tablenotemark{a}	& $\alpha$	& $\delta$	& $\Delta$D$_{assoc}$		\\
		& 							& (J2000)		& (J2000)	& (kpc)				 	 \\
\tableline
1	& M82		& 09$^h$50$^m$07.2$^s$	& 69$^{\circ}$55$^`$56$^{``}$	& 31				   \\
2	& NGC3077	& 10$^h$02$^m$50.7$^s$	& 68$^{\circ}$19$^`$49$^{``}$	& 23				 \\		   
3	& NGC2976	& 09$^h$52$^m$44.9$^s$	& 68$^{\circ}$12$^`$50$^{``}$	& 27					 \\
4	& M81		& 10$^h$01$^m$45.0$^s$	& 69$^{\circ}$16$^`$31$^{``}$	& 33  				 \\
5	& M82		& 09$^h$55$^m$06.3$^s$	& 69$^{\circ}$22$^`$05$^{``}$	& 18 				\\   
\tableline
\tablenotetext{a}{Most likely association, based on positional and velocity proximity}

\end{tabular}
\end{center}
\end{table}

\begin{table}
\begin{center}
\caption{\hone Cloud Detections: Derived Properties}
\begin{tabular}{l|c|c|c|c|c|c}
\tableline\tableline
Cloud	& Peak Flux Density		& Integrated Flux	& V$_{HI}$\tablenotemark{a}	& $\sigma_{V}$\tablenotemark{a,b} 	& $\vert$ V$_{HI}$-V$_{assoc}$ $\vert$	& Mass	\\
		& (Jy)				& (Jy km s$^{-1}$)	& (km s$^{-1}$) 			& (km s$^{-1}$)					& (km s$^{-1}$)						& ($\times 10^7 M_{\odot}$)	     		\\ \tableline
1	& 0.11 $\pm$ 0.08			& 0.91 $\pm$ 0.08	& 165					& 50							& 38 	       							& 1.47 $\pm$ 0.35\\	
2	& 0.10 $\pm$ 0.08			& 1.39 $\pm$ 0.09	& -105		& 55		& 119  	       	& 2.25 $\pm$ 0.49	\\
3	& 0.12 $\pm$ 0.08			& 1.65 $\pm$ 0.10	& 11			& 82		& 8	   	       	& 2.67 $\pm$ 0.65\\
4	& 0.30 $\pm$ 0.08			& 5.18 $\pm$ 0.28	& 72			& 28	     	& 106  	       	& 8.37 $\pm$ 1.75\\
5	& 0.07 $\pm$ 0.08			& 0.43 $\pm$ 0.08	& 280		& 36	     	& 77   	       	& 0.69 $\pm$ 0.27\\
\tableline
\tablenotetext{a}{From Gaussian fit to \hone profile}
\tablenotetext{b}{FWHM calculated from Gaussian fit}
\end{tabular}
\end{center}
\end{table}

\end{document}